\documentclass[twocolumn,twocolappendix]{aastex7}
\usepackage{amsmath}
\usepackage{graphicx}
\usepackage{soul}
\usepackage{hyperref}

\newcommand{\ramses}{\textsc{ramses}}
\newcommand{\bpass}{\textsc{bpass}}
\newcommand{\ramsesrt}{\textsc{ramses-rt}}

\definecolor{red}{RGB}{250,0,0}
\definecolor{magenta}{RGB}{167,0,135}

\definecolor{pink}{RGB}{255,51,153}

\definecolor{green}{RGB}{0,153,0}

\definecolor{blue}{RGB}{0,0,250}

\definecolor{grey}{RGB}{180,180,180}

\setstcolor{red}

\newcommand{\myr}{\,{\rm Myr}}
\newcommand{\gbzero}{\texttt{gB0}}
\newcommand{\gbone}{\texttt{gB1}}

\newcommand{\gbzerowhd}{\texttt{gB0\_wHD}}
\newcommand{\gbonewhd}{\texttt{gB1\_wHD}}

\newcommand{\gbzerowmhd}{\texttt{gB0\_wMHD}}
\newcommand{\gbonewmhd}{\texttt{gB1\_wMHD}}

\newcommand{\whd}{\texttt{wHD}}
\newcommand{\wmhd}{\texttt{wMHD}}

\newcommand{\mg}{\mbox{$\mu$G}}

\newcommand{\msun}{\mbox{$\rm M_{\odot}$}}
\newcommand{\msunyr}{\mbox{$\rm M_{\odot}\,{\rm yr^{-1}}$}}
\newcommand{\kms}{\mbox{${\rm km\,s^{-1}}$}}

\newcommand{\cmq}{\mbox{$\,{\rm cm^{-3}}$}}
\newcommand{\nH}{\mbox{$n_{\rm H}$}}

\defcitealias{lee20}{L20}
\defcitealias{lee22}{L22}

\received{\today}
\revised{\today}

\shorttitle{Jellyfish Galaxies in Magnetic Fields}
\shortauthors{Jaehyun Lee et al.}

\begin{document}

\title{Jellyfish Galaxies in Magnetic Fields: Insights from Numerical Simulations}

\author[0000-0002-6810-1778]{Jaehyun Lee}
\affiliation{Korea Astronomy and Space Science Institute, 776, Daedeokdae-ro, Yuseong-gu, Daejeon 34055, Republic of Korea}
\affiliation{Korea Institute for Advanced Study, 85 Hoegi-ro, Dongdaemun-gu, Seoul 02455, Republic of Korea}
\email[show]{syncphy@gmail.com}

\author[0000-0002-3950-3997]{Taysun Kimm}
\affiliation{Department of Astronomy, Yonsei University, 50 Yonsei-ro, Seodaemun-gu, Seoul 03722, Republic of Korea}
\email[show]{tkimm@yonsei.ac.kr}

\author[0000-0002-7534-8314]{J\'er\'emy Blaizot}
\affiliation{Univ Lyon, Univ Lyon1, Ens de Lyon, CNRS, Centre de Recherche Astrophysique de Lyon UMR5574, F-69230 Saint-Genis-Laval, France}
\email{jeremy.blaizot@univ-lyon1.fr}

\author[0000-0002-8140-0422]{Julien Devriendt}
\affiliation{Astrophysics, University of Oxford, Denys Wilkinson Building, Keble Road, Oxford OX1 3RH, UK}
\email{julien.devriendt@physics.ox.ac.uk}

\author[0000-0002-4059-9850]{Sergio Martin-Alvarez}
\affiliation{Kavli Institute for Particle Astrophysics \& Cosmology (KIPAC), Stanford University, Stanford, CA 94305, USA}
\email{martin-alvarez@stanford.edu}

\author[0000-0002-0184-9589]{Jinsu Rhee}
\affiliation{Department of Astronomy, Yonsei University, 50 Yonsei-ro, Seodaemun-gu, Seoul 03722, Republic of Korea}
\affiliation{Korea Astronomy and Space Science Institute, 776, Daedeokdae-ro, Yuseong-gu, Daejeon 34055, Republic of Korea}
\email{jinsu.rhee@gmail.com}

\author[0009-0007-7943-0378]{Maxime Rey}
\affiliation{Department of Astronomy, Yonsei University, 50 Yonsei-ro, Seodaemun-gu, Seoul 03722, Republic of Korea}
\email{maximerey.astro@gmail.com}

\author{Adrianne Slyz}
\affiliation{Astrophysics, University of Oxford, Denys Wilkinson Building, Keble Road, Oxford OX1 3RH, UK}
\email{adrianne.slyz@physics.ox.ac.uk}

\begin{abstract}
Jellyfish galaxies provide direct evidence of ram pressure stripping in cluster environments. We investigate the role of magnetic fields in the formation of jellyfish galaxies with a multiphase interstellar medium (ISM) using radiation magneto-hydrodynamic simulations. We impose magnetized (MHD) and non-magnetized (HD) winds on the gas-rich dwarf galaxies containing the magnetized or non-magnetized ISM. The MHD winds strip the disk gas more effectively than the HD winds because of the magnetic force acting against the local density gradient, which results in remarkably different ram pressure stripped features. The magnetic fields induced by the MHD winds generate a strong magnetic pressure, which forms smoothed disks and tail gas features. Since the stripped ISM in MHD wind cases travels while being nearly isolated from the intracluster medium (ICM), the stripped ISM mostly forms stars within 20~kpc of the galactic disks. In contrast, non-magnetized winds facilitate the efficient mixing of the stripped ISM with the ICM, resulting in the formation of abundant warm clouds that cool and collapse in the distant ($\sim50-100\,$kpc) tails at times of a few hundred Myr. Consequently, distant tail star formation occurs only in the HD wind runs. Finally, despite the different tail features, the star formation rates in the disk remain similar owing to the interplay between the increased gas stripping and the gas density increase in the disks of the MHD wind runs. These results suggest that the magnetized ICM may have a significant influence on jellyfish galaxies, whereas the magnetized ISM play a minor role.
\end{abstract}


\section{Introduction}

Ram pressure stripping has a significant impact on galaxies in galaxy clusters because of high peculiar velocity and high ambient medium density~\citep{gunn72}. Ram pressure stripped (RPS) tails are observed mostly in cluster environments in a wide wavelength range from HI~\citep{kenney04,oosterloo05,chung07,chung09,scott10,scott12,scott18}, radio continuum~\citep{gavazzi95,chen20,ignesti21,roberts21},  CO~\citep{jachym14,verdugo15,jachym17,lee17,lee18,moretti18,jachym19}, H$\alpha$~\citep{gavazzi01,cortese06,cortese07,sun07,yagi07,yagi10,fumagalli14,Boselli16,poggianti17, sheen17}, to X-ray bands~\citep{finoguenov04,wang04,machacek05,sun05,sun06,sun10}, indicating the multiphase nature of the RPS tails. \citet{sun21} find a characteristic flux ratio of X-ray to H$\alpha$ from the RPS tails, which is interpreted as evidence of mixing between the RPS clouds and the intracluster medium (ICM). \citet[][ hereafter L22]{lee22} demonstrate using a radiation hydrodynamical (RHD) simulation for a gas-rich dwarf galaxy that the flux ratio closely correlates with the fraction of stripped interstellar medium (ISM) in the RPS tails. In this RHD simulation, dense molecular clumps form in the distant tails as a significant amount of ISM gas is stripped off by the strong ram pressure and mixes with the ICM, resulting in a lower metallicity for molecular clumps or stars that are more distant from the galactic plane. A metallicity gradient of the gas clumps is also found in the RPS tails formed by varying the ram pressure~\citep[see also][]{sparre24}. This is consistent with the metallicity gradient observed in the RPS tails~\citep{franchetto21}. 

Magnetic fields are ubiquitous in galaxies and galaxy clusters~\citep{carilli02,govoni04,beck13, beck19a, Lopez-Rodriguez22b} even at high redshifts~\citep{Bernet08, Lopez-Rodriguez22b}.
Spiral galaxies typically have total fields of $|\mathbf{B}|\lesssim10\,$\mg~\citep{niklas95, beck15}, bright spirals with well-developed arms have $|\mathbf{B}|\sim10-20\,$\mg~\citep{fletcher11}, and starburst galaxies have a magnetic field strength of $|\mathbf{B}|\sim50-300\,$\mg~\citep{lacki13,adebahr13,adebahr17} with magnetized outflows \citep{Lopez-Rodriguez21}. Observations for synchrotron relic and halo radio sources~\citep[e.g.,][]{ensslin98,brunetti01,arlen12,degasperin17,vanweeren19,Cuciti22,Botteon22}, inverse Compton X-ray emission~\citep[e.g.,][]{rapaeli99,ge20}, Faraday rotation of polarized radio sources~\citep[e.g.,][]{vogt05,bonafede10,kuchar11,bohringer16,govoni17,osinga22,Xu22}, and cluster cold fronts in X-ray~\citep[e.g.,][]{vikhlinin01} reveal that the ICM is magnetized with a typical field strength of $\sim1-10\mu$G, depending on the location in the clusters and the topology of the magnetic fields.

Magnetic fields influence the ISM and circumgalactic medium (CGM) in complicated ways. As an important contributor to the internal energy budget, magnetic pressure is assumed to disturb the gravitational collapse of clouds, possibly lowering star-formation activities~\citep{federrath12,kauffmann13,Zamora-Aviles_18}. Besides, aligned magnetic fields stabilize flows, thereby suppressing the development of dynamical instabilities, namely the Rayleigh-Taylor instability~\citep{jun95}, the nonlinear thin shell instability~\citep{vishniac94,heitsch07}, and the Kelvin-Helmholtz instability~\citep{chandrasekhar61}, which in turn interrupt the efficient mixing of two different media~\citep[e.g.,][]{frank96,ryu00,esquivel06,heitsch09,hamlin13,liu18,praturi19}. When magnetic fields are perpendicular to gas flows, they reduce gas cooling and thermal instability~\citep{field65}, which in turn suppresses cloud fragmentation and the buildup of density contrast~\citep{burkert04,heitsch08b,heitsch08a}. 
Therefore, magnetic fields have a significant impact on stripped clouds in the CGM \citep[e.g.,][]{jennings23,sparre24a}

Because the ICM is typically magnetized with $|\mathbf{B}|\sim1-10\mu$G in $R\lesssim1\,$Mpc~\citep[e.g.,][]{carilli02} and galaxies also have magnetic fields, it is not surprising to observe magnetized RPS tails in clusters. The presence of magnetic fields has been confirmed in several RPS galaxies~\citep{gavazzi95,vollmer10,chen20,ignesti21,muller21b,muller21a,muller21c}. \citet{gavazzi95} estimate that the tails of their two target galaxies have a magnetic field strength of $|\mathbf{B}|\sim5$\mg\ based on equipartition arguments \citep[see][for potential overestimation of magnetic fields when the equipartition assumption is applied.]{dacunha24}. \citet{muller21a} measure the magnetic field orientation and strength of the well-studied RPS galaxy JO206. They show that the tail of JO206 is not highly turbulent, its field strength is $|\mathbf{B}|>4.1\,$\mg, and the field direction is aligned with the tail. Based on other studies on the multiphase tails of JO206 and JW100, \citet{muller21b} propose a scenario in which magnetic fields produce a magnetized interface in the tails that protects RPS clouds from the hot ICM.

Several studies have investigated the effect of magnetic fields on RPS galaxies in cluster environments using magneto-hydrodynamical (MHD) simulations~\citep[e.g.][]{ruszkowski14,tonnesen14,vijayaraghavan17,ramos-martinez18,sparre24a}. \citet{ruszkowski14} conducts a set of simulations in which both MHD and pure hydrodynamical (HD) winds are imposed on non-magnetized disks. They find that MHD winds form a magnetic draping layer~\citep[see][for further details]{dursi08} that lowers the gas stripping rate in the disk and renders the tail clouds more filamentary than in the HD winds. \citet{tonnesen14} examine the behavior of MHD and HD disks when the HD winds are blowing toward the disks. In these experiments, the stripped clouds mix with the ICM less efficiently in the case of MHD wind. \citet{vijayaraghavan17} find that the magnetized tails are smoother and narrower than the tails of their pure HD counterparts~\citep{vijayaraghavan15}. \citet{ramos-martinez18} demonstrate that gas inflow is induced from the outer disk toward the central region when an MHD disk is exposed to HD wind. As in previous studies, the RPS tails are less clumpy in the magnetized disk case. \citet{sparre24a} demonstrate that jellyfish galaxies tend to have magnetic fields aligned with their RPS tails. They also find a stronger star formation boost from galaxies encountering winds in the edge-on direction than those encountering winds in the face-on direction \citep[see also][]{akerman23}. However, these studies are mostly based on their own idealized conditions with no {\it in situ} star formation, and hence no stellar feedback driving the strong turbulent motion on gaseous disks. Turbulent pressure contributes significantly to the internal energy of clouds~\citep[e.g.][]{joung06,kim13,choudhury21,ostriker22}, providing additional resilience along with a gravitational restoring force against the ram pressure~\citep[][]{lee20,choi22}. Furthermore, as described above, magnetic fields suppress the mixing between different media. In \citetalias{lee22}, we present the role of mixing in the formation of the characteristic RPS tails using RHD simulations. However, we do not include a magnetized medium in the models, which calls for further studies on the formation process of jellyfish galaxies in the conditions of a magnetized ISM or ICM. In this study, we perform radiation magneto-hydrodynamical (RMHD) simulations with rich physical ingredients, to better understand the formation process of jellyfish galaxies.

The remainder of this paper is organized as follows. Section~\ref{sec:simulations} describes the RMHD method and initial conditions of our simulations. In Section~\ref{sec:galactic_disk}, the impact of ram pressure on gas disks and star formation activities in both the RMHD simulations and their RHD counterparts are examined. Section~\ref{sec:rps_tails} demonstrates the different properties of RPS tails in the RMHD and RHD cases. In Section~\ref{sec:discussion}, we discuss the results and present several caveats. Finally, Section~\ref{sec:conclusions} summarizes the major findings.

\section{Simulations}
\label{sec:simulations}
In this section, we describe the code and initial conditions of the simulated galaxies and the properties of the ICM winds in the simulations.

\subsection{Code}
\label{sec:code}
We conduct a suite of RMHD simulations using the adaptive mesh refinement (AMR) code \ramsesrt~\citep{teyssier02,rosdahl13,rosdahl15a}. To solve the induction equation for modeling the evolution of magnetic fields in the AMR framework of \ramses, \citet{fromang06} implement an extended MUSCL-Hancook scheme into \ramses, based on the constrained transport algorithm. \ramsesrt\ adopts the approximate Harten-Lax-van Leer contact wave~\citep{toro94} to solve the set of Euler equations for the MHD fluids and the particle-mesh method~\citep{guillet11} to solve the Poisson equation. Radiative transfer (RT) is computed using a momentum-based scheme with the Global Lax-Friedrich flux function. {We use a modified version of RAMSES-RT that self-consistently follows the photoionization and photodissociation of hydrogen and helium species. The code tracks eight photon groups, spanning energies from 0.1–1.0 eV (infrared; relevant for non-thermal radiation pressure) to $>$54.42 eV (extreme ultraviolet; capable of ionizing HeII). The photon groups also include the Lyman–Werner band (11.2–13.6 eV), which plays a crucial role in computing} the formation and destruction of molecular hydrogen using a modified photochemistry model~\citep{katz17,kimm17}, on top of the non-equilibrium chemistry and cooling of six chemical species, HI, HII, HeI, HeII, HeIII, and $e^-$.

\ramses\ constructs an AMR structure based on a fully threaded tree scheme~\citep{khokhlov98}. In this study, a gas cell is refined when the local thermal Jeans length is smaller than 32 cells until the maximum refinement level is attained. This refinement scheme enables us to intensively resolve most of the RPS tail volumes with a maximum or second maximum grid level. More details can be found at the end of Section~\ref{sec:ICM_Winds}.

 \begin{figure*}
\centering 
\includegraphics[width=0.85\linewidth]{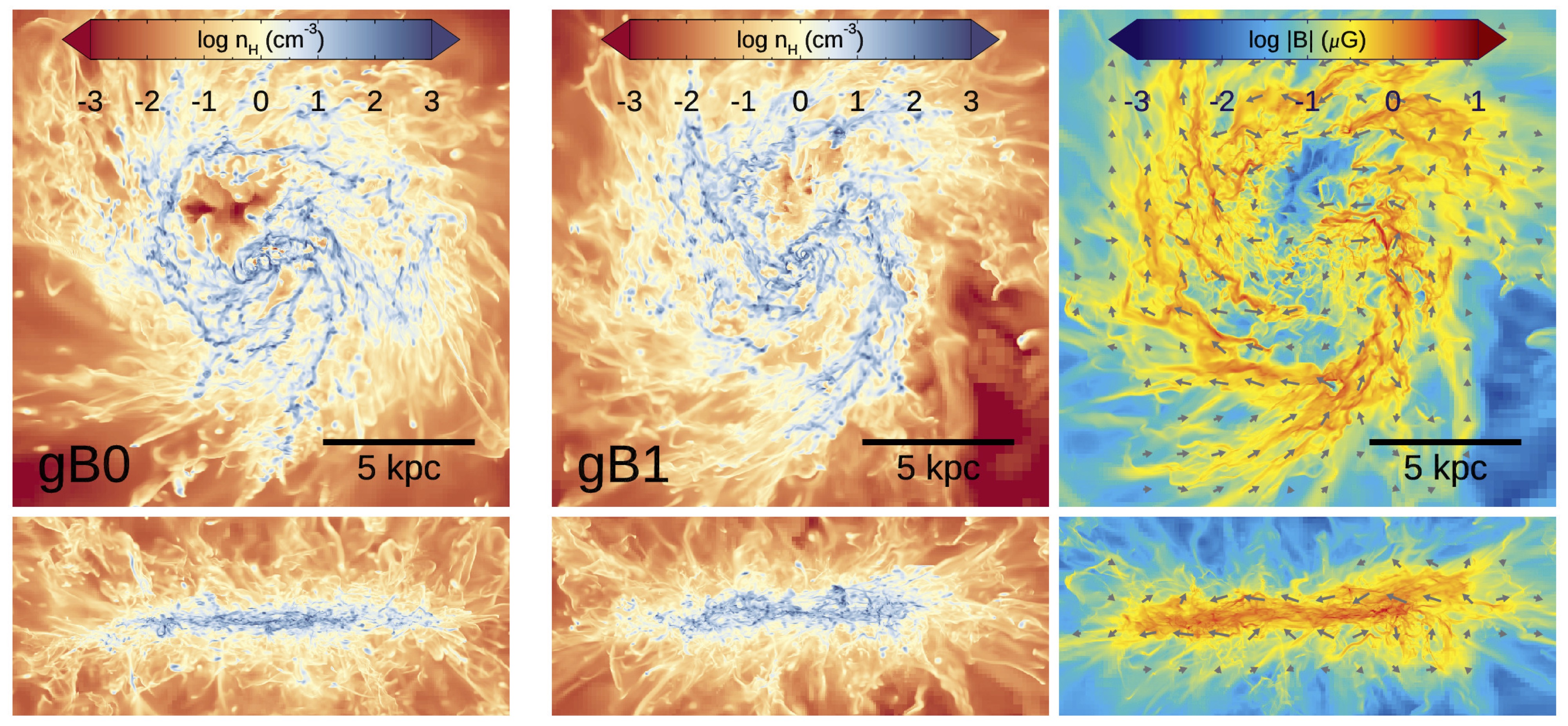}

\caption{Maps of the gas density and magnetic field strength projected in the face-on and edge-on directions at $t=100\,$Myr, when winds start to influence the galaxies in simulations with ICM winds. Grey arrows show the direction of the magnetic fields and their relative field strength averaged over $\Delta z=6\,{\rm kpc}$ on a logarithmic scale.} 
\label{fig:ic}
\end{figure*}

Atomic metal cooling at $T\gtrsim10^4\,{\rm K}$ is computed using the Cloudy cooling model~\citep{ferland98} and fine structure line cooling at $T\lesssim10^4$K using the cooling model of \citet{rosen95}. In our simulations, we also include radiative cooling induced by molecular hydrogen~\citep{hollenbach79,halle13}. The spectral energy distributions of stars computed using the Binary Population and Spectral Synthesis model~\cite[\bpass\ version 2.0,][]{eldridge08,stanway16} based on a Kroupa initial mass function (IMF) is used with slopes of -1.3 for stellar masses between 0.1 and $0.5\,\msun$, and -2.35 for stellar masses between 0.5 and $100\msun$ \citep{kroupa01}.

Star formation rates (SFRs) are computed based on the Schmidt law~\citep{schmidt59}:
\begin{equation}
\label{eqn:schmidt}
\frac{{\rm d}\rho_{\rm star}}{{\rm d}t}=\epsilon_{\rm ff}\frac{\rho_{\rm gas}}{t_{\rm ff}},
\end{equation}
where $\rho_{\rm gas}$ is the gas density, $t_{\rm ff}=\sqrt{(3\pi/32G\rho_{\rm gas})}$ is the free fall time, $G$ is the gravitational constant, and $\epsilon_{\rm ff}$ is the star formation efficiency per free-fall time.
In our model, the star formation efficiency $\epsilon_{\rm ff}$ is assumed to be controlled by the local thermo-turbulent state of gas~\citep{padoan11,federrath12}. This is implemented in our model as follows~\citep[e.g.,][]{kimm17}:
\begin{equation}
\label{eqn:eff}
\epsilon_{\rm ff}=\frac{\epsilon_{\rm ecc}}{2\phi_{\rm t}} \exp \bigg( \frac{3}{8} \sigma^2_s \bigg) \bigg[ 2- {\rm erf} \bigg( \frac{\sigma^2_s-s_{\rm crit}}{\sqrt{2\sigma^2_s}}\bigg) \bigg],
\end{equation}
where $\epsilon_{\rm ecc}\approx0.5$ is the fraction of gas in a cloud that can collapse into stars without being affected by proto-stellar jets, expanding shells, or outflows, $1/\phi_{\rm t}\approx0.47$ is a numerical factor that parameterizes the multi-free-fall time in the cloud. $\sigma^2_s=\ln [1+b^2\mathcal{M}^2\beta/(1+\beta)]$ is the standard deviation of the logarithmic density contrast $s\equiv\ln(\rho/\rho_0)$, where $b=0.4$ represents a mode of turbulence driving and corresponds to a mixture of solenoidal and compressive turbulence. $\mathcal{M}$ is the sonic Mach number, $\beta$ is the ratio of thermal pressure to magnetic pressure ($P_{\rm th}/P_{\rm B}$), and $\rho_0$ is the mean density of the cloud~\citep[see e.g.][for a detailed derivation]{molina12}. The critical density above which the gas may start to collapse ($s_{\rm crit}$) is approximated as \citep{federrath12}
\begin{equation}
\label{eqn:scrit}
s_{\rm crit}=\ln[0.067\theta^{-2}\alpha_{\rm vir}\mathcal{M}^2f(\beta)],
\end{equation}
where $\theta=1$ is a numerical factor for the uncertainty in the post-shock thickness with respect to the cloud size, $\alpha_{\rm vir}\equiv2E_{\rm kin}/|E_{\rm grav}|$ is the virial parameter of the cloud, and $f(\beta)\equiv(1+0.925\,\beta^{-1.5})^{2/3}/(1+\beta^{-1})^2$. In a hydrodynamic case, the magnetic pressure $P_{B}=0$ makes plasma $\beta$ diverge to infinity, $f(\beta)$ and $\beta/(1+\beta)$ converge to 1. The local Mach number is estimated for a cell from the velocity dispersion between neighburing cells as $\sigma_{\rm gas}^2={\rm Tr} \left( \vec{\nabla} v^{\rm T} \vec{\nabla} v \right) \Delta x^2$, where $v$ is the divergence and rotation-free local velocity field. The typical $\epsilon_{\rm ff}$ in our simulations is $\sim$0.2--0.3. Once the efficiency is determined, a Poisson distribution is used to randomly sample the mass of the star particle with the minimum mass of $M_{\rm star,min}=914\,\msun$, following \citet{rasera06}. In our simulations, SFRs are computed only in cells with $n_{\rm H}>100\,{\rm cm^{-3}}$. We confirm that stellar particles mostly ($\sim97.5\%$) form in gas cells with hydrogen number density $n_{\rm H}\gtrsim300\,{\rm cm^{-3}}$ because star formation is only enabled in gravitationally self-bound clouds in our star formation scheme.

\begin{figure}
\centering 
\includegraphics[width=\linewidth]{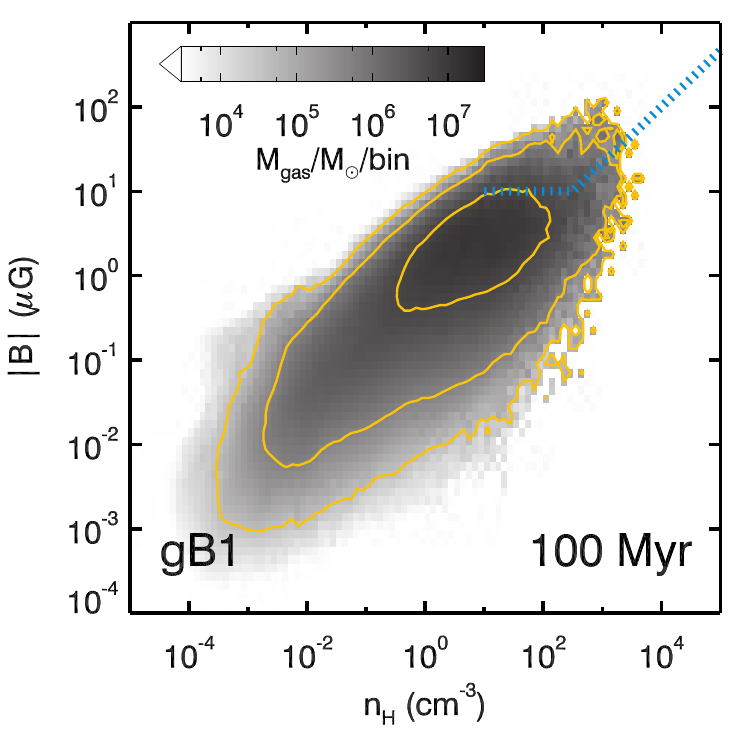}
\caption{Relation between the gas density and the magnetic field strength in the disk plane of the \gbone\ galaxy at $t=100\,$Myr, when winds start to influence the galaxies. The grey shades display disk gas mass distribution{ and the yellow contours denote $M_{\rm gas}=10^{4.5}$, $10^{5.5}$, and $10^{6.5}\,\msun/{\rm bin}$ from outside to inside.}. The blue dotted lines show the most probable maximum values of the ISM magnetic field strength at densities $n_{\rm H}>10\,{\rm cm^{-3}}$~\citep{crutcher10}.  }
\label{fig:bmag_rho}
\end{figure}

{ Various stellar feedback mechanisms are considered, i.e., photoionization, radiation pressure exerted by photons at wavelengths ranging from the Lyman continuum to optical~\citep{rosdahl13}, and non-thermal radiation pressure owing to IR \citep{rosdahl15a} and Lyman alpha photons \citep{kimm18}, based on on-the-fly radiation transport.} Type II supernova (SN) explosions are modeled using the mechanical feedback scheme of \citet{kimm15}, with the SN frequency increased by a factor of five compared to the canonical value to provide strong support against gravitational collapse in the ISM. Note that such an increase is necessary to reproduce the stellar mass growth rates and the UV luminosity functions of galaxies at $z\ge 6$ \citep[e.g.,][]{rosdahl18,garel21} or the mass fraction of stars in galaxies similar to the Milky Way~\citep{li18}\footnote{\bf We note that although the Milky Way and high-redshift gas-rich galaxies differ substantially in their global properties, both systems appear to require a similarly enhanced level of feedback to reproduce the galactic properties. Without it, simulated galaxies are overly compact and too rapidly rotating, as already shown by \citet{kimm15}.}. We do not allow metal enrichment from SNe to trace the origin of the gas in a cloud. {Note that star formation in the tail is governed by freefall collapse and is unlikely to be affected by the neglect of self-enrichment, because the freefall timescale of the warm ionized clouds ($n_{\rm H} \sim 0.1,{\rm cm}^{-3}$) is much longer ($\sim 100$ Myr) than their cooling timescale~\citepalias[$\sim10\,$Myr, see][for details]{lee22}.} { AGN feedback is not included in this study since no clear or strong evidence of a significant link between AGN and star formation activities in low mass ($M_\star\lesssim3\times10^{10}\,\msun$) galaxies has been found~\citep{martin-navarro18}.} Interested readers can refer to \citetalias{lee20} for more details on radiation hydrodynamics.

\subsection{Initial Conditions}
We adopt the initial condition of a gas-rich dwarf galaxy employed in \citetalias{lee22}. The initial conditions are generated by \citet{rosdahl15b} using the \texttt{MAKEDISK} code~\citep{springel05a}. The simulation box has a side length of $300\,$kpc, covered by $256^3$ root grids (i.e., level 8). The grids are refined until they reach a maximum refinement level of 14 which corresponds to 18\,pc. The dwarf galaxy is placed at the center of a dark matter halo with total mass $M_{\rm halo}=10^{11}\,\msun$ and radius $R_{\rm vir}=89\,$kpc. {The dark matter halo is composed of dark matter particles of mass of $M_{\rm DM}=10^5\,\msun$ and it} is generated using an NFW density profile~\citep{navarro97} with a concentration parameter $c=10$ and a spin parameter $\lambda=0.04$. The dwarf galaxy initially has a stellar mass of $M_\star=2.1\times10^9\,\msun$ and a HI mass of $M_{\rm HI}=8.4\times10^9\,\msun$. The HI disk is set to have a metallicity of $Z_{\rm ISM}=0.75\,Z_{\odot}$, where $Z_\odot=0.0134$~\citep{asplund09}. {We do not place the circumgalactic medium (CGM) around the model galaxy because the CGM can be quickly stripped even before entering the virial radius of a galaxy cluster~\citep[e.g.][]{bahe13,zinger18}.}

A galactic disk is placed on the $xy$ plane of the simulation box. A magnetic field strength of $B_x=0.1\mu G$ (or 0 if unmagnetized) is initially imposed on the gaseous disk. The initial magnetic field is amplified to a typical observed strength of a few $\mg$ after compression of the gaseous disk. The simulations for the two disks are denoted as \gbzero\ and \gbone, respectively. The galaxies enter a quasi-equilibrium state {in the mass-weighted PDF of gas thermal pressure} after $\sim150\,$Myr, which is defined as $t=0$. 

Figure~\ref{fig:ic} shows the projected distribution of the hydrogen number density and the magnetic field strength of the two galaxies with initial $B_x=0$ (left, \gbzero) and $B_x=0.1\mu G$ (middle and right, \gbone) at $t=100\,\myr$. There is no significant difference in the gas density distributions of the two galaxies in either projection. Figure~\ref{fig:bmag_rho} displays the magnetic field strength as a function of the hydrogen number density of the \gbone\ galaxy. After relaxation ($\Delta t=250\,{\rm Myr}$), the disk gas mass of the \gbone\ galaxy is mostly distributed in $|\mathbf{B}|\sim1-10\,\mu$G in $n_{\rm H}>10\,{\rm cm^{-3}}$, which reasonably agrees with the empirical relation of \citet[][blue dotted line]{crutcher10} or the typical magnetic field strength of the disk galaxies~\citep{beck15}. 

As radiation and momentum feedback from young stars exert outward pressure on the ISM, it has been theorized that the sum of the turbulent, thermal, and magnetic pressures in a galaxy closely correlates with the SFR \citep[e.g.,][]{ostriker10,ostriker22}. 
We also find that our simulated galaxies follow the total pressure--SFR relation of \citet{ostriker22} (See \ref{sec:SFR_pressure} for more details), which indicates that they are in a reasonable pressure equilibrium.

 \begin{deluxetable}{lccccc}
\tablewidth{0.85\linewidth}
\tabletypesize{\scriptsize}
    \tablecaption{The initial parameters for the six galaxies modelled in this paper. The galaxy starts with a stellar mass of $M_\star=2.1\times10^9\,\msun$ and is embedded in a dark matter halo of $M_{\rm halo}=10^{11}\,\msun$. The simulations are set to have a spatial resolution up to $\Delta x=18\,$pc (level 14). From left to right, each column indicates the model name, the ICM density ($n_{\rm H,\,ICM}$), the ICM velocity ($v_{\rm ICM}$), wind ram pressure ($P_{\rm ram}/k_{\rm B}$), the initial disk ($\mathbf{B}_{\rm disk}$) and the wind ($\mathbf{B}_{\rm ICM}$) magnetic field strength.}
        \label{tab:ic}

    \tablehead{
    \colhead{Model}  & \colhead{$n_{\rm H,\,ICM}$} & \colhead{$v_{\rm ICM}$} & \colhead{ $P_{\rm ram}/k_{\rm B}$} & \colhead{$\mathbf{B}_{\rm disk,0}$} & \colhead{$\mathbf{B}_{\rm ICM}$}\\ 
      \colhead{} & \colhead{[${\cmq}$]} & \colhead{ [$\kms$]}  & \colhead{[${\rm cm^{-3}\,K}$]}  & \colhead{[$10^{-6}G$]} & \colhead{[$10^{-6}G$]} 
       }
       \startdata
     \gbzero & $10^{-6}$ & 0 & 0 &  0 & - \\
      \gbone & $10^{-6}$ & 0 & 0   &  $B_x=0.1$  & -\\ 
      \hline
     \gbzerowhd & $3\times10^{-3}$&$10^3$ & $5\times 10^5$ & 0 & 0 \\
      \gbonewhd & $3\times10^{-3}$&$10^3$ &  $5\times 10^5$   &  $B_x=0.1$  & 0\\ 
            \hline
     \gbzerowmhd & $3\times10^{-3}$&$10^3$ & $5\times 10^5$ &  0 & $B_x=1$ \\
      \gbonewmhd & $3\times10^{-3}$&$10^3$ &  $5\times 10^5$   &  $B_x=0.1$  & $B_x=1$\\ 
    \enddata
 
\end{deluxetable}

\subsection{ICM Winds}
\label{sec:ICM_Winds}
 
We impose two ICM winds on the galaxies {in the face-on direction} at $t=0$. Motivated by the observations of the nearby clusters~\citep[e.g.,][]{tormen04,hudson10,urban17}, the two winds commonly have $v=1000\,{\rm km\,s^{-1}}$, $n_{\rm H}=3\times10^{-3}\,{\rm cm^{-3}}$, $T=3\times10^7\,K$, and $Z_{\rm ICM}=0.3\,Z_{\odot}$, however, the HD wind case (\texttt{wHD}) has no magnetic fields, while the other is magnetized with $B_x=1\,$\mg\ (\texttt{wMHD}). The magnetic field strength adopted in this study is comparable to that inferred from observations of cluster environments.~\citep{govoni04,vogt05,bonafede10}. However, we note that the magnetic field strength can vary significantly between clusters or even within a cluster depending on the topology of the magnetic field~\citep{carilli02,kuchar11}. The winds exert a ram pressure of $P_{\rm ram}/k_{\rm B}=5\times10^5\,{\rm K\,cm^{-3}}$ on the simulated galaxies, where $k_{\rm B}$ is the Boltzmann constant. The ram pressure mimics the typical ram pressure that a galaxy may experience in a cluster center (see Figure~10 of \citealt{jung18} {or Figure~6 of \citealt{yun19}}). {Since the sound speed of the ICM is $c_s=880\,\kms$, the wind velocity relative to the galaxy is supersonic ($\mathcal{M}=1.14$), consistent with the results found in the IllustrisTNG100 simulation~\citep{yun19}.} As a control sample, galaxies in the absence of wind is evolved as well.


With an initial velocity of $v_{\rm ICM}=1000\,\kms$, the wind front arrives at the galaxies at $t\approx 135\,$Myr. However, the winds begin to affect the galaxies slightly earlier than their arrival time because the winds move forward, pushing the medium in front of them owing to shock expansion. All the simulations end at $t=380\,$Myr from the wind launch or at 530\,Myr from the start of the simulations. Each run takes $\sim1.4\,$M CPU hours with 640 cores when the winds are imposed. 

We define the galactic disk as the components enclosed by the cylindrical volume of radius $r=12\,$kpc and height $h=\pm3\,$kpc from the galactic mid-plane. Therefore, the RPS tail is defined as a structure enclosed in a long cylinder with $r=12\,$kpc and $h>3\,$kpc. Table~\ref{tab:ic} lists the six simulations performed in this study and their setups. Because of our intensive refinement scheme, $33-45\%$ of the volume of the RPS tails is filled with the maximum level grids at the epoch when the ICM winds cross the simulation boxes ($t=300\,$Myr). The volume-filling fraction increases to $62-85\%$ when the second maximum level grids are included. This indicates that our simulations sufficiently resolve not only the ISM in the disks but also the stripped clouds in the CGM.

\section{Impact of Ram Pressure on Galactic Disks}
\label{sec:galactic_disk}

\begin{figure*}   
\centering 
\includegraphics[width=0.85\linewidth]{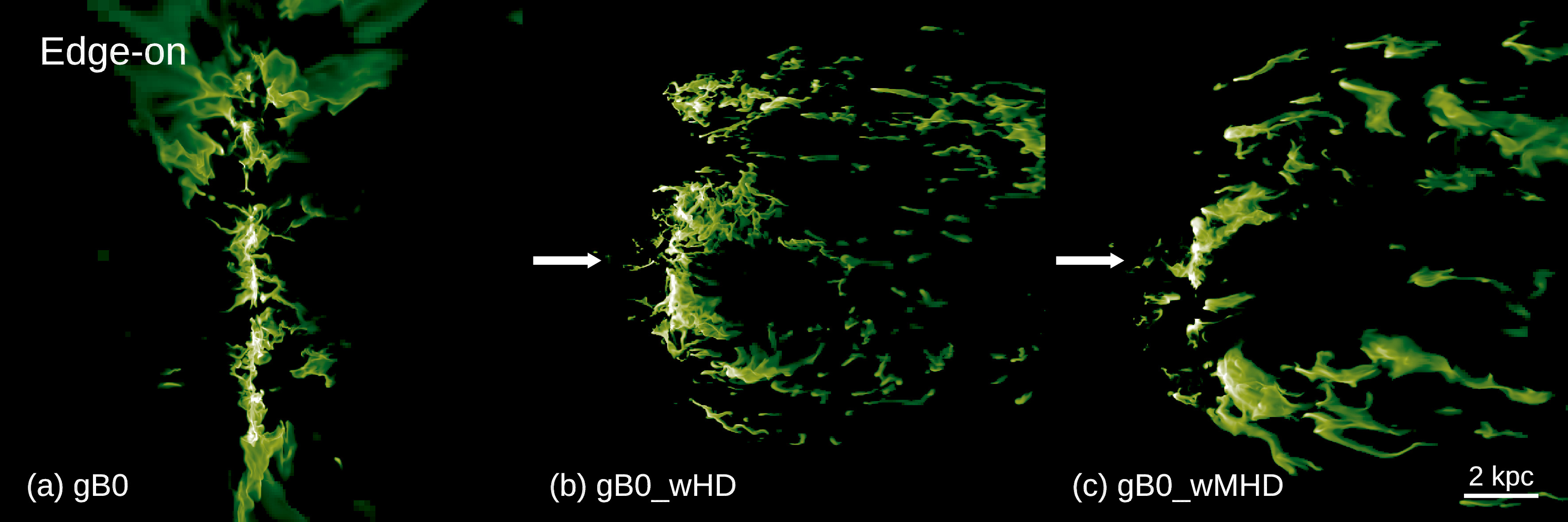}
\includegraphics[width=0.85\linewidth]{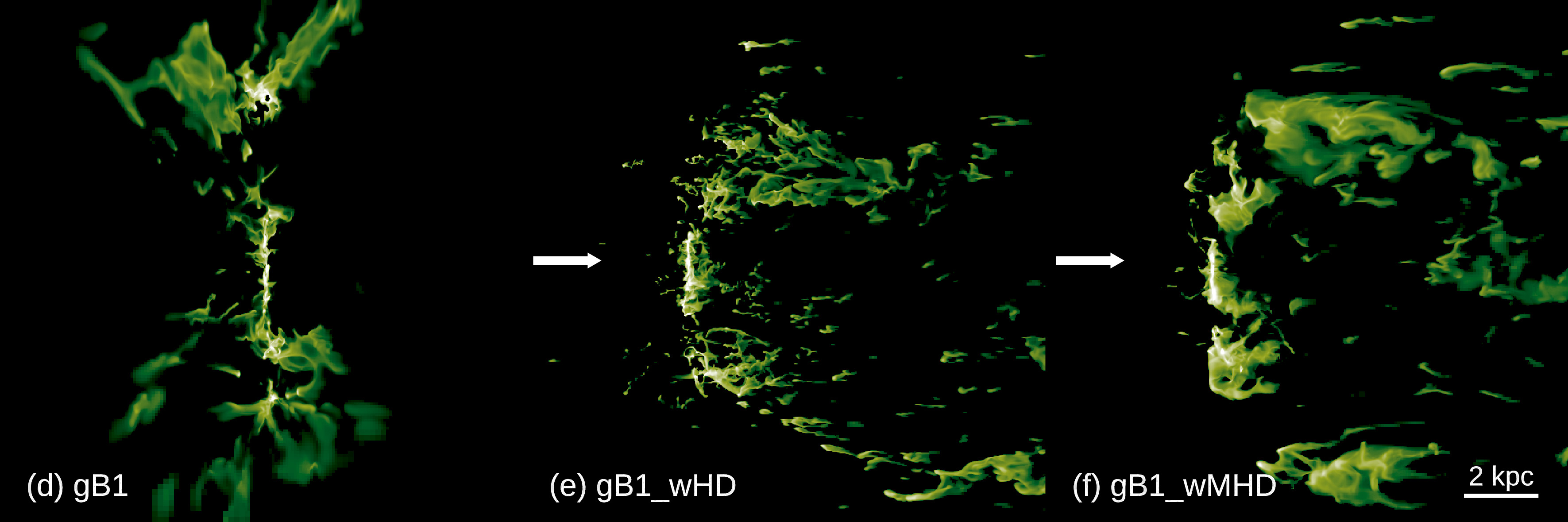}
\includegraphics[width=0.85\linewidth]{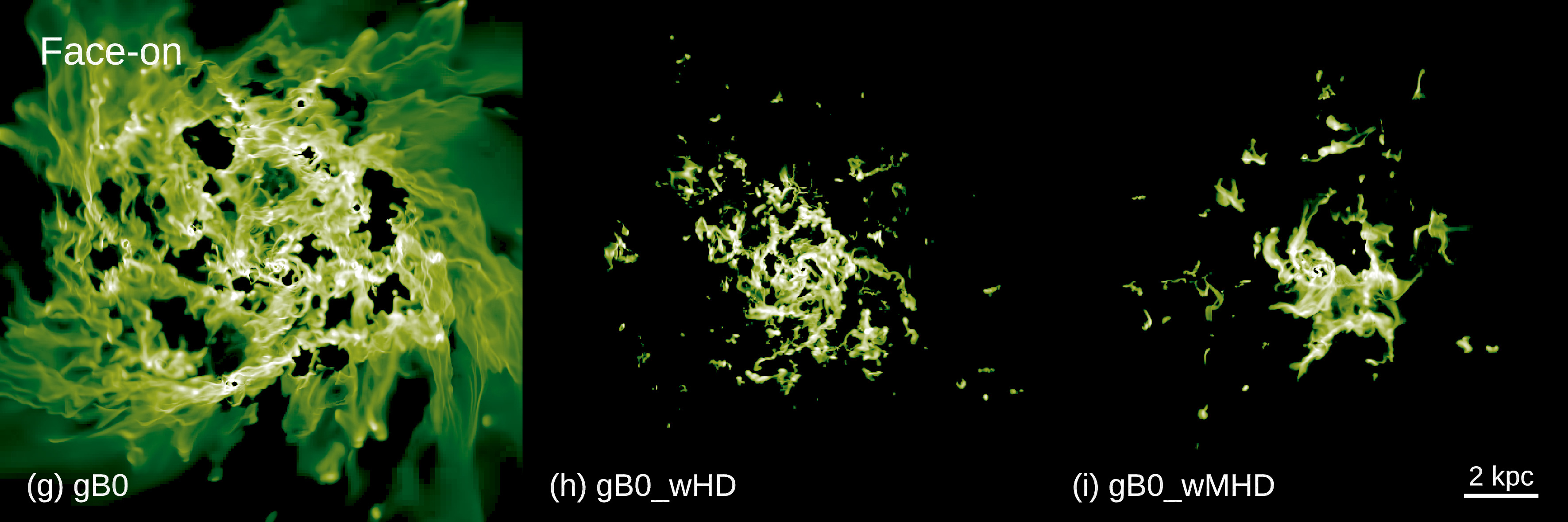}
\includegraphics[width=0.85\linewidth]{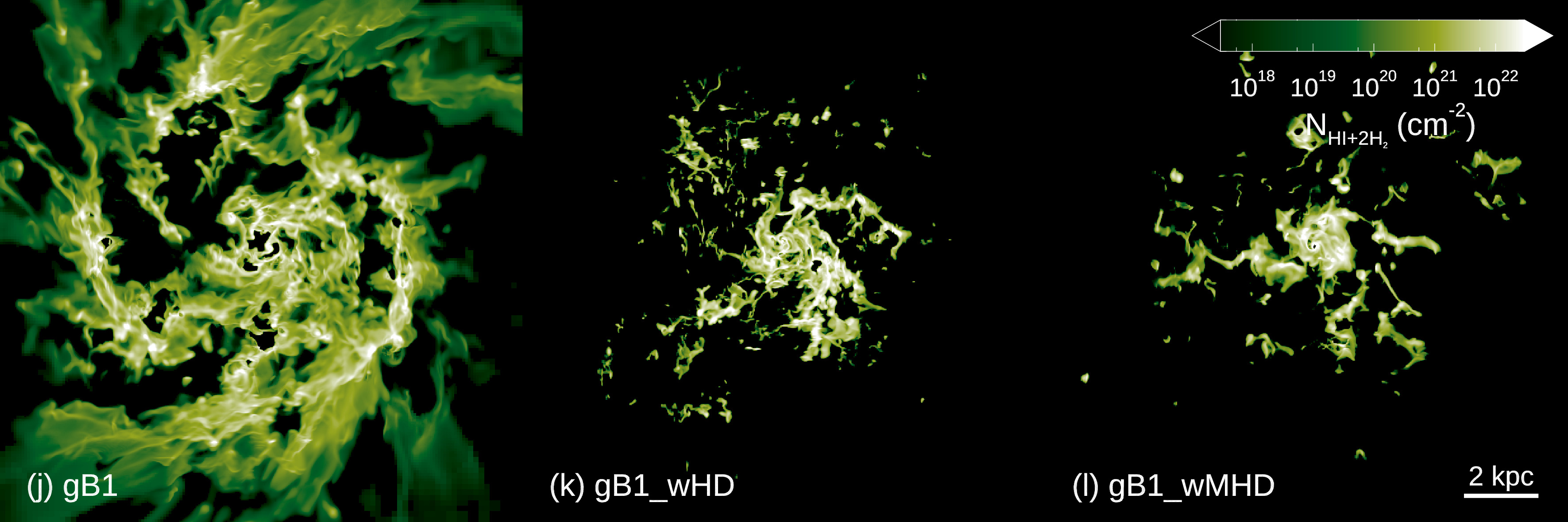}
\caption{Edge-on (upper two rows) and face-on (bottom two rows) cold gas (HI+H$_2$) column density at $t=150~$Myr, right after the winds encounter the disks. We integrate it over a slab of $\Delta y=500$pc, centered at the center of the disk stellar mass. White arrows indicate the direction of winds in the edge-on projections. The cool gas disk and the stripped tails are more clumpy with HD winds (\whd) than with MHD winds (\wmhd) because magnetised winds smooth the gas while stripping it from the galaxy.
}
\label{fig:disk_projection}
\end{figure*}

The ICM winds imposed in our simulations are designed to mimic the strong ram pressure exerted in the central regions of massive galaxy clusters \citep[e.g.,][]{jung18}. Such strong ram pressure quickly removes most of the ISM and suppresses disk star formation activity within $\sim200\,{\rm Myr}$, as shown in \citetalias{lee22}. Extending upon \citetalias{lee22}, we first examine the impact of ram pressure on galactic disks in the presence of magnetic fields, and then discuss its effect on the RPS tails in the next section.

\subsection{Overall Evolution of the RPS Disk in the Presence of Magnetic Fields}
\label{sec:disk_stripping}

Previous studies have suggested that gaseous disks are less prone to stripping when the ISM is magnetized~\citep{tonnesen14,ramos-martinez18}. \citet{ruszkowski14} also demonstrated that a magnetized ICM creates magnetic draping layers that hinder  stripping of the ISM from the disk. However, it is important to note that these studies are based on an idealized setup without star formation and stellar feedback, and the trend may be different in a more realistic setting, as shown in this section.

\begin{figure}
\centering 
\includegraphics[width=0.95\linewidth]{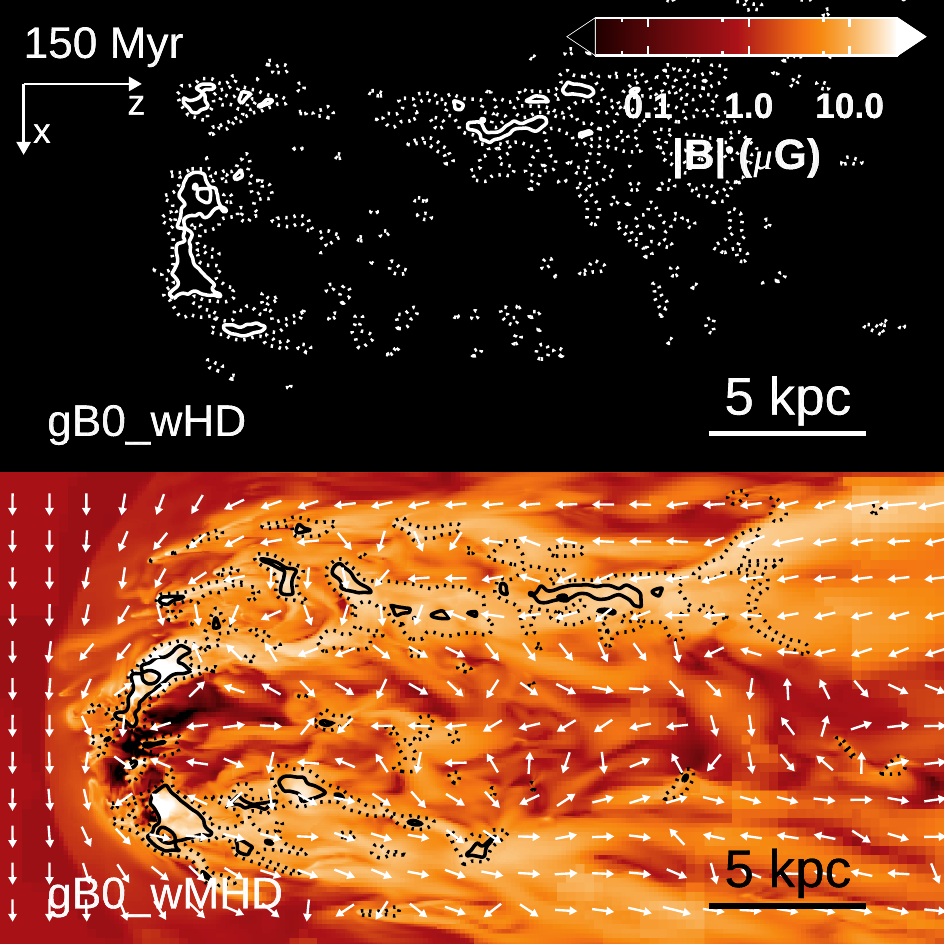}
\includegraphics[width=0.95\linewidth]{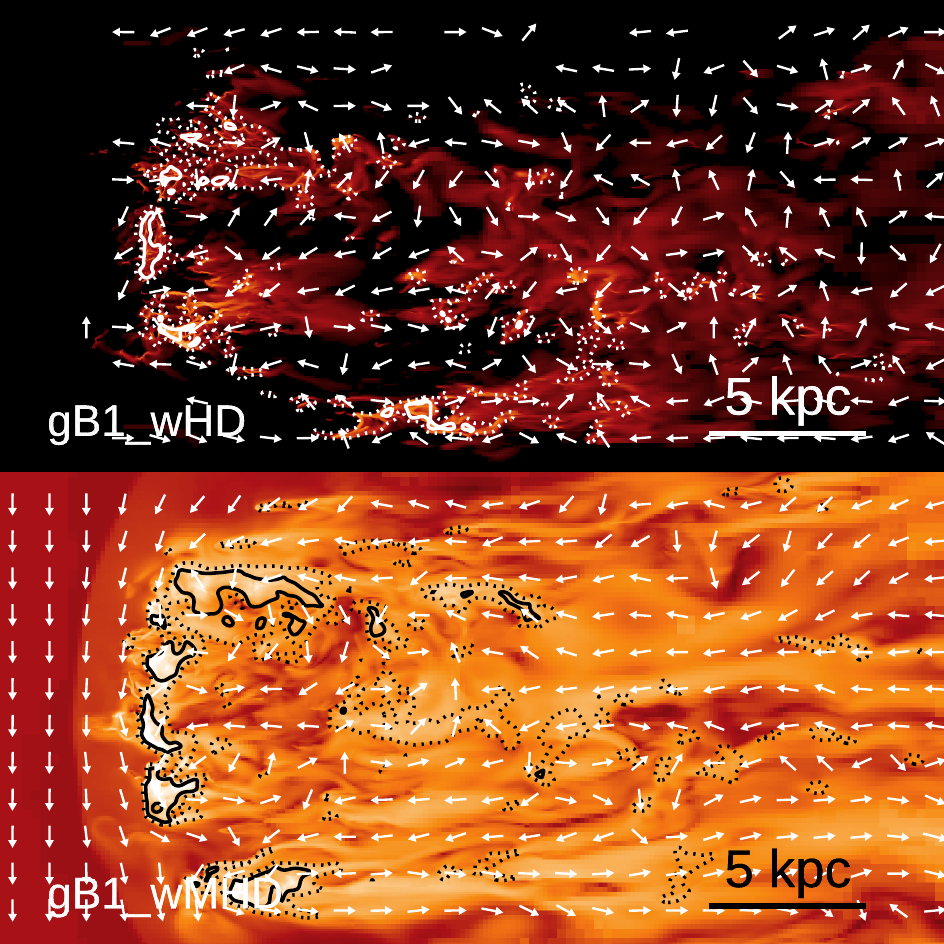}
\caption{Projected magnetic field strength in the central slab ($-250\leq y \leq250 {\rm pc}$ from the center of the stellar mass) of the runs including winds at $t=150\,$Myr. The magnetic field strength is volume-weighted. Dotted and solid contours denote the projected density of cold (HI+H$_2$) gas $N_{\rm H}=10^{16}$ and $10^{20}\,{\rm cm^{-2}}$, respectively. For visibility, the contours are colored white for the \whd\ runs and black for the \wmhd\ runs. White arrows trace the magnetic field orientation in the plane of the projection for regions with $|\mathbf{B}|\ge10^{-2}\,\mg$. The field orientation is averaged in a volume of $1.17\times1.17\times0.5\, {\rm kpc^3}$, for visibility. 
The winds blow perpendicular to the galactic disk. The simulations with magnetized ICM winds (\wmhd) show a stronger field in an order of magnitude in the disks owing to magnetic draping. }
\label{fig:disk_bfield}
\end{figure}

Figure~\ref{fig:disk_projection} shows the cold gas (HI+H$_2$) column density of the RPS disks measured on a slab with a thickness of 500~pc crossing the galactic center in the edge-on (two upper rows) and face-on (two lower rows) directions. The first notable feature is that the stripped clouds are less fragmented/shattered in the runs with magnetized winds (right columns, \wmhd) than in the HD wind cases (middle columns, \whd). Similar to the stripped gas, the galactic disks struck by magnetized wind are smoother and less fragmented because both magnetic draping and shock compression amplify the magnetic fields in the disk~\citep{sparre20}. 
The amplification of the disk and tail magnetic fields is clearly shown in Figure~\ref{fig:disk_bfield}, in which we plot the magnetic field strength on a central slab with a thickness 500~pc for different simulations. While the magnetized disk hit by the HD wind shows $|\mathbf{B}|\sim1\,\mu$G (\gbzerowhd), the magnetic fields are enhanced by more than an order of magnitude when the disk is exposed to the magnetized ICM wind (\gbzerowmhd\ and \gbonewmhd, see Sec.~\ref{sec:alignment_amplification} for further discussion on the amplification in the tail). However, in contrast to the magnetic fields in the wind, the presence of pre-existing disk magnetic fields does not drastically alter the clumpy features in the galactic disk and stripped gas (e.g., \gbzero\ and \gbone\ in the panels (g) and (j) of Figure~\ref{fig:disk_projection}), although strong disk fields can in principle mildly smooth the ISM structures~\citep[e.g.,][]{kortgen19,martin-alvarez20}.

\begin{figure}
\centering 
\includegraphics[width=\linewidth]{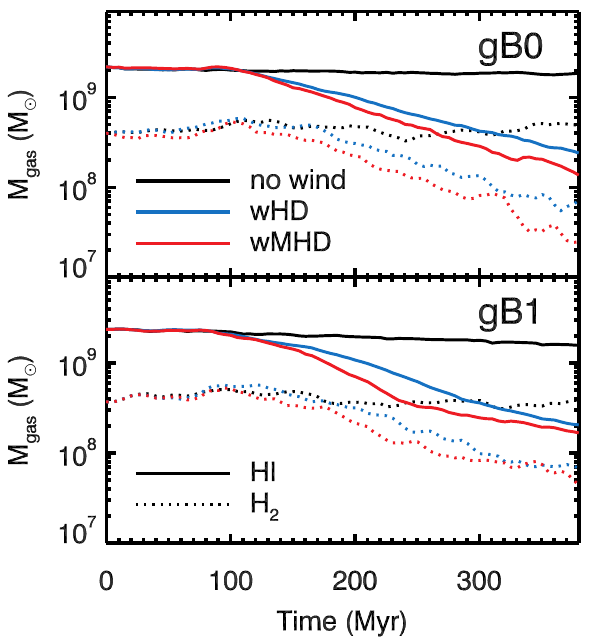}
\caption{HI (solid) and H$_2$ (dotted) mass evolution in the disk of gB0 (top) and gB1 (bottom) without winds (black), with HD winds (blue) and with MHD winds (red). Disk gas is commonly stripped more by the MHD winds than the HD winds.}
\label{fig:gas_disk}
\end{figure}

Figure~\ref{fig:gas_disk} shows the masses of HI (solid) and H$_2$ (dotted) in the galactic disk as a function of time. In the absence of ram pressure (black), the \gbzero\ and \gbone\ disks have similar HI and H$_2$ masses, respectively, with less than 10\% difference before the arrival of the wind ($t\lesssim100\,$Myr). These masses are rapidly reduced to less than half within $\sim 100\,{\rm Myr}$ when the strong ICM winds are imposed \citepalias[see also][]{lee22}. HI stripping is slightly more efficient in the \gbonewhd\ galaxy than in the \gbzerowhd\ galaxy; however, the difference is not significant. After careful examination of the vertical structure just before the ICM--ISM interaction, we conclude that the more efficient stripping in \gbonewhd\ is likely owing to a burst of star formation that coincidentally reduces the column density. Although it is possible that the additional magnetic support in the disk acts against gravity, which results in more effective stripping, this is unlikely to be the case here because the \gbonewhd\ galaxy is preferentially supported by turbulent pressure (see below) and not by magnetic pressure. Therefore, it is difficult to conclude that the magnetized disks (\gbonewhd) are more susceptible to stripping than the non-magnetized disks (\gbzerowhd).

More interestingly, we find that magnetized ICM winds (\wmhd) remove more gas than the HD winds (\whd), as the strong magnetic fields induced by magnetic draping permeate and inflate the ISM (see Sec.~\ref{sec:density}), helping the ICM wind strip the gas. 
This contrasts with previous studies that report less efficient gas stripping as a result of the development of magnetic draping layers. By performing additional simulations without gas cooling and star formation (see \ref{sec:draping}), we confirm the previous finding that disk stripping is less efficient in a disk with magnetic draping layers when the disk is smooth and not turbulent. Effective ISM stripping with magnetized winds is discussed in more detail below and in Section \ref{sec:51}.

\begin{figure}
\centering 
\includegraphics[width=\linewidth]{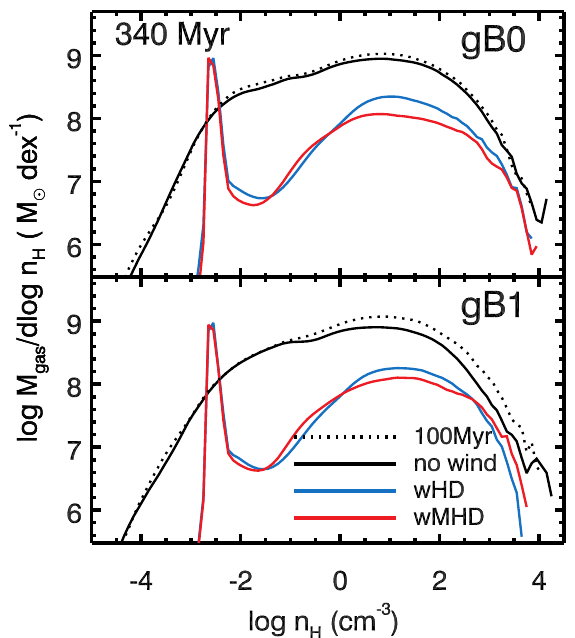}
\caption{Disk gas density functions of the simulated galaxies at $t=100\,$Myr (dotted lines) and $t=340\,$Myr (solid lines) without winds (black), with HD winds (blue) and with MHD winds (red). We do not show the gas density for the cases with winds at $t=100\,$Myr as they are similar to those without winds (dotted black line). Without ram pressure, the gas density functions do not significantly evolve after $t=100\,$Myr. The strong ram pressure destructively strips the disk gas in regions with densities $n_{\rm H}\lesssim10^{3}\,{\rm cm^{-3}}$. At densities $n_{\rm H}\sim10^{0}-10^{2}\,{\rm cm^{-3}}$, more gas is stripped when the ICM is magnetized (red). The sharp peaks at $n_{\rm H}\sim10^{-2.5}\,{\rm cm^{-3}}$ are formed by the ICM winds enclosed in the disk regions.}
\label{fig:disk_gas_pdf}
\end{figure}

\begin{figure}
\centering 
\includegraphics[width=\linewidth]{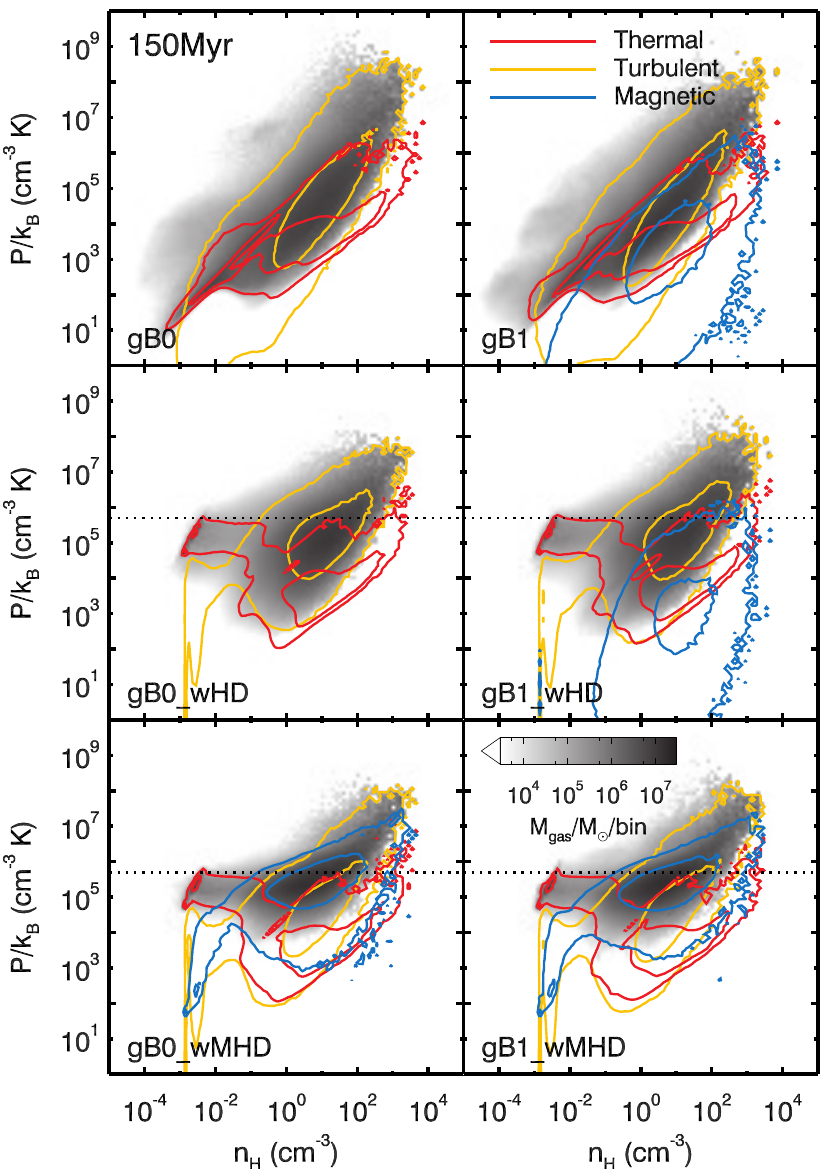}
\caption{Contribution of different components to the total gas pressure in the disk at $t=150\,$Myr, just before the gaseous disks are largely stripped. Different color codes indicate three pressure components, i.e., thermal (red), turbulent (yellow), and magnetic (blue) pressure. The horizontal dotted lines indicate the ram pressure. The inner and outer contours indicate the distribution of bins with masses of $3\times10^7$ and $10^5\,\msun/{\rm bin}$, respectively. The magnetic pressure becomes as significant as the turbulent pressure in the disks encountering MHD winds. The red or blue contour clumps seen at $n_{\rm H}\sim10^{-3.5}\,{\rm cm^{-3}}$ below the horizontal dotted line are formed by the ICM enclosed within the disk volume.}
\label{fig:pressure_rho_two}
\end{figure}

\begin{figure}
\centering 
\includegraphics[width=0.9\linewidth]{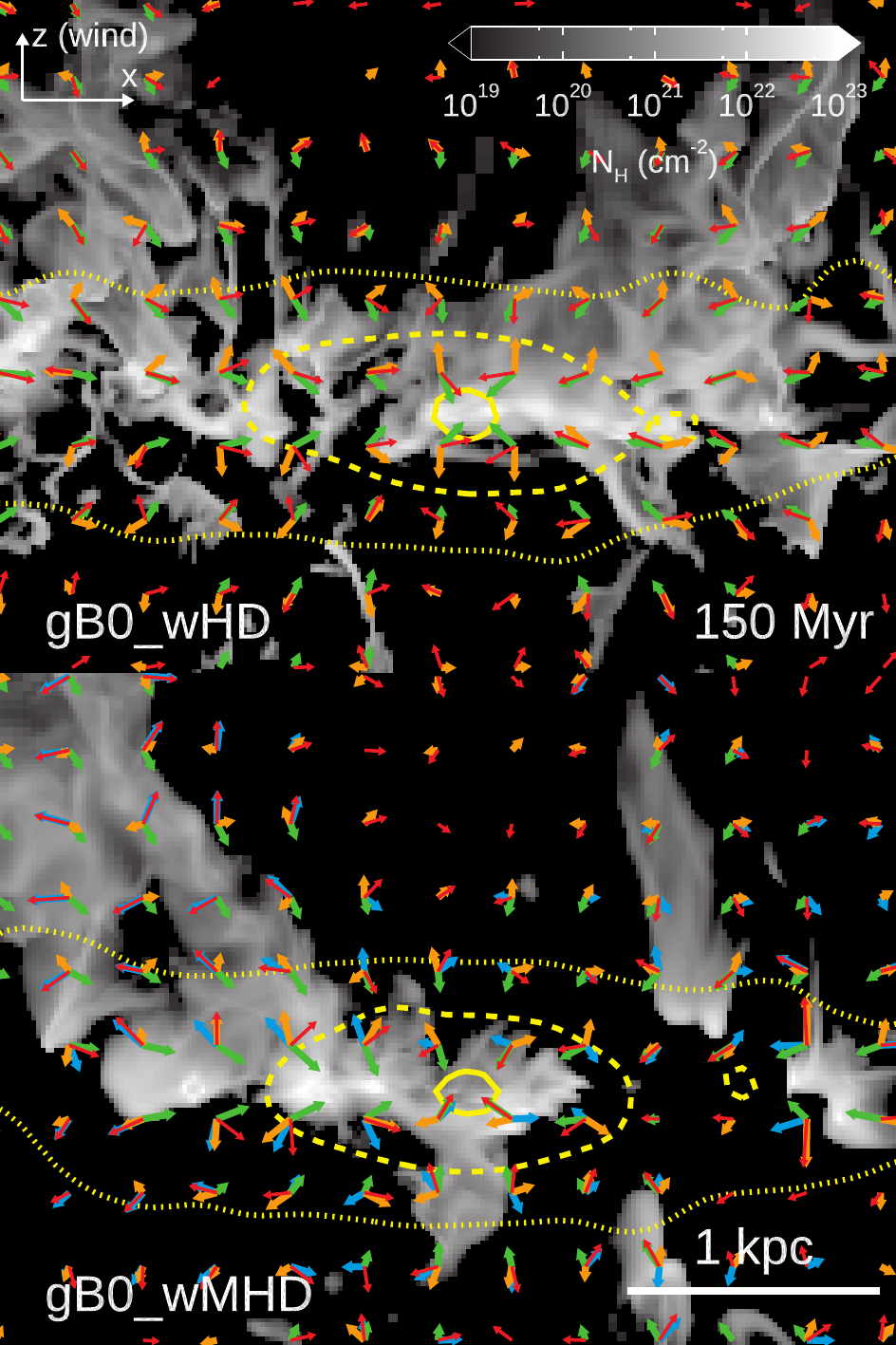}
\caption{Projected force density fields in the central slab ($-250\le y \le 250$pc from the center of the stellar mass) of the \gbzero\ galaxy encountering the HD (top) and MHD (bottom) winds {in the edge-on views of the disks}. The grey shades show the column density of cold (HI+H$_2$) gas. The {yellow} contours denote the stellar column density of  $2\times10^2\,M_{\odot}\,{\rm pc^{-2}}$ (dotted), $6\times10^2\,M_{\odot}\,{\rm pc^{-2}}$ (dashed), and $2\times10^3\,M_{\odot}\,{\rm pc^{-2}}$ (solid). The red, {orange,} blue, and green arrows indicate the direction of net, {turbulent,} magnetic, and gravitational forces. The net force is the vector sum of magnetic, turbulent, thermal, and gravitational forces. The length of the arrows mark the force density strength in a logarithmic scale. The force fields are plotted when the force density $f>10^{-33}{\,\rm dyn\,cm^{-3}}$. In \gbzerowmhd, the magnetic fields exert an additional force that is significantly stronger than the local gravity in stripped clouds while gravity is weak but dominant in the stripped wakes of \gbzerowhd. This is also the case for \gbone.}
\label{fig:force_field}
\end{figure}

\begin{figure*}
\centering 
\includegraphics[width=\linewidth]{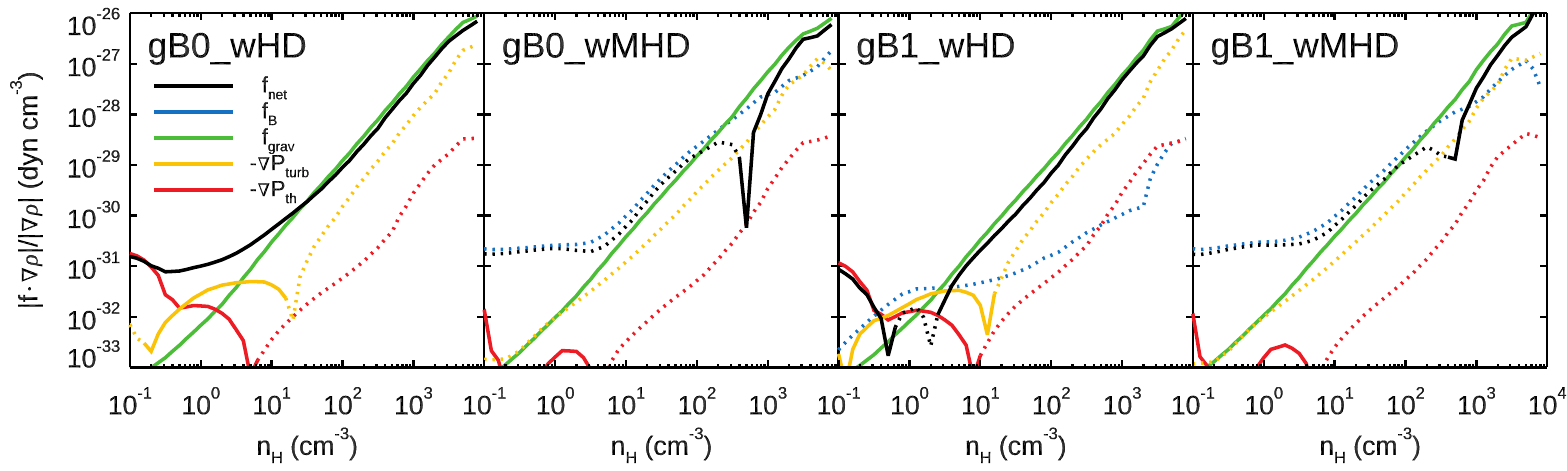}
    \caption{Strength of the force density aligned with the gas density field gradient as a function of the hydrogen number density. The blue, green, yellow, and red lines indicate the magnetic force, gravity, and the forces induced by turbulent pressure and thermal pressure, respectively. $f_{\rm net}$ (black) is the sum of the four force components. Dotted lines show the negative inner product between the forces and the density gradient. The force densities are volume-weighted median, averaged over $t=150\pm20\,$Myr. The net force acts against gravity at $n_{\rm H}\lesssim100\,{\rm cm^{-3}}$ owing to the strong magnetic force when the ICM is magnetized.}
\label{fig:f_rho}
\end{figure*}

\subsection{Effect of Magnetized Winds on Gas Density Distributions}
\label{sec:density}
The smooth structure of the galactic disk in the \wmhd\ runs indicates that the \gbzerowmhd\ and \gbonewmhd\ galaxies may have different gas density distributions in the disk compared to those encountering non-magnetized winds. Figure~\ref{fig:disk_gas_pdf} shows the gas density distribution just before the disks encounter winds (black dotted, $t=100\pm40\,$Myr) or after the disks are sufficiently stripped (solid, $t=340\pm40\,$Myr). 
A comparison between the black dotted and solid lines shows that the density distributions do not change significantly between $t=100$ and 340$\,$Myr in the absence of the ICM wind. However, after the arrival of the ICM wind, both galaxies lose large amounts of gas over the entire density range. As expected, the gas loss is particularly pronounced at lower densities. The peak at $n_{\rm H}\sim10^{-2.5}\,{\rm cm^{-3}}$ is the result of the ICM filling the disk regions after stripping.

We find that the intermediate density gas with $n_{\rm H}\sim10^{0}-10^{2.5}\,{\rm cm^{-3}}$ is stripped more efficiently in the \wmhd\ runs (red) than in the \whd\ runs (blue). This is consistently observed in both \gbzero\ and \gbone\ throughout the time after the arrival of the winds. This also explains the significant gas stripping in the magnetized wind cases shown in Figure~\ref{fig:gas_disk}. 
We attribute this to the complex interplay between the strong magnetic and turbulent support and the ram pressure. Figure~\ref{fig:pressure_rho_two} shows the contributions of the turbulent (yellow), thermal (red), and magnetic (blue) pressures to the total pressure budget (sum of the three components) in the disk immediately after the wind front arrives at the galaxies. The turbulent pressure is computed as $P_{\rm turb}=\rho_{\rm gas}\,\sigma^2_{\rm gas}$, where $\sigma_{\rm gas}^2$ is the same as that in Section~\ref{sec:code} for estimating the local Mach number. For comparison, the ram pressure is also plotted through the ICM as a dotted line ($P_{\rm ram}$). We show that in the \wmhd\ runs (third row) the magnetic pressure dominates over the turbulent or thermal pressure in the range of $n_{\rm H}\sim10^{0}-10^{2}\,{\rm cm^{-3}}$. Given that a strong magnetic pressure can counteract the ram pressure, it may be surprising that the ISM is stripped more in the \wmhd\ runs.
However, the magnetic pressure can also act against the gravitational collapse of clouds, allowing more ISM in the diffuse phase, as shown in Figure~\ref{fig:disk_projection} \citep[see also][]{hennebelle13,kortgen19,martin-alvarez20,robinson24}. 
Indeed, we confirm that the gaseous disks, once magnetized by the MHD winds, have consistently larger scale heights \footnote{Here the disk scale height is computed as $H=\sqrt{\int\rho z^2{\rm d}V/\int\rho {\rm d}V}$~\citep{kim13} within the half-mass radius of a gaseous disk and $|z|\le 3\,$kpc, where $\rho$ is the density of a cell, $z$ is the vertical distance from the galactic midplane to the cell, and $dV$ is the volume of the cell.} within their half-mass radii than the disks in the HD wind runs. At $t=150\pm25\,$Myr, just before the disks are largely stripped by the winds, the disks of \gbzerowhd\ and \gbonewhd\ have $H\sim0.82\,$kpc, while the disks of \gbzerowmhd\ and \gbonewmhd\ have $H\sim0.96\,$kpc. 
As the ram pressure is still greater than the magnetic pressure of the disks of the \wmhd\ runs in the density range and as the momentum is continuously transferred from the ICM to the ISM~\citep{choi22}, the intermediate density gas eventually becomes susceptible to stripping in \gbzerowmhd\ and \gbonewmhd, increasing the low density components ($10^{-1.5}\,\cmq \la \nH \la 10^{0}\,\cmq$) in Figure~\ref{fig:disk_gas_pdf}.

To further elucidate the role of magnetized winds in enhancing gas stripping, we analyze the interplay between gravity, magnetic fields, turbulent pressure, and thermal pressure in the RPS disks and tails. Figure~\ref{fig:force_field} illustrates the force density fields within a $\Delta y=500\,$pc-thick slab centered on the stellar mass distribution. The net force (red arrows) is the sum of the four components: turbulent (orange arrows), thermal, gravitational (green arrows), and magnetic (blue arrows) forces. We remove rotational and laminar velocity fields from local flows to measure turbulent motion of fluids. Therefore, the flow of the ICM is not presented in the turbulent force field. The magnetic force is the sum of magnetic tension and magnetic pressure terms. Gravitational forces predominantly act toward the galactic center. Within the galactic disk, the turbulent pressure driven by the star formation processes is a dominant force, competing closely with gravity. In \gbzerowmhd\ (bottom), the net force field is predominantly influenced by a strong magnetic force on the outskirts of the disk or stripped clouds, whereas it is weaker and tends to align with gravity when the ICM wind is unmagnetized (top). This suggests that magnetic fields play a crucial role in shaping the distinct cloud features between the \whd\ and \wmhd\ runs. 

The effect of the force components is quantified by associating the force fields with the density fields. Figure~\ref{fig:f_rho} shows the strength of the force density fields aligned with the gradient of the gas density field ($|f\cdot\nabla\rho|/|\nabla\rho|$) as a function of gas density, where $f$ is the volume-weighted median force density and $\rho$ is the gas density in a cell. The gravitational force density $f_{\rm grav}$ always has monotonically increasing positive values with increasing gas density, whereas the volume-weighted magnetic field force density $f_{\rm B}$ is always negative, which indicates that the magnetic field tends to inflate the gas against gravity. This is not surprising because magnetic fields are generally stronger in denser clouds, thereby creating outward magnetic pressure. {Thermal pressure also produces a force in the opposite direction to the density gradients. The force induced by turbulent pressure is also negative, and is particularly strong in dense clouds. This is because turbulent pressure is mainly driven by strong stellar feedback in galactic disks.}
When the ICM is magnetized (\wmhd), $f_{\rm B}$ dominates the other forces at $n_{\rm H}\lesssim100\,{\rm cm^{-3}}$, and eventually forms the negative net force opposite to the gas density gradient. This accounts for the smooth features of the disk and stripped clouds in the runs with MHD winds.



\begin{figure}
\centering 
\includegraphics[width=1\linewidth]{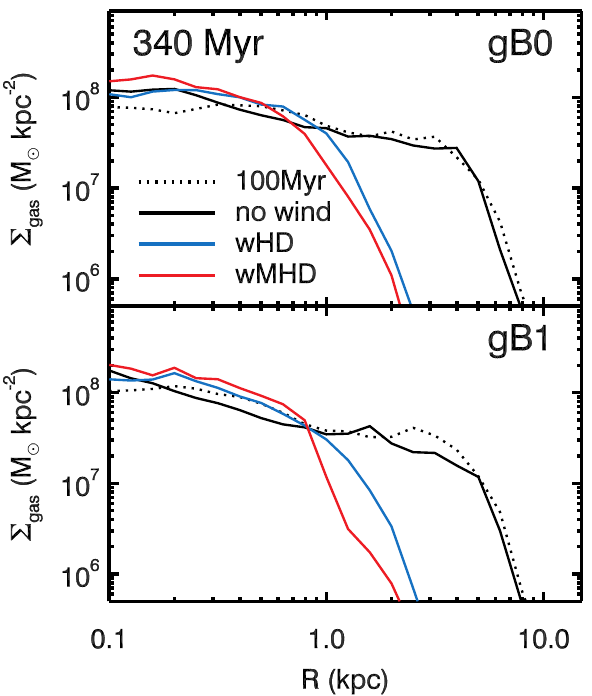}
\caption{Gas column density profiles in the disk region as a function of the cylindrical radius ($R$) of \gbzero\ and \gbone\ averaged over $\pm40\,$Myr centered at $t=100\,$Myr (dotted), $t=340\,$Myr (solid). As before, we show the simulation without winds in black, with HD winds in blue, and with MHD winds in red. We do not show the HD and MHD wind cases at $t\sim100\,$Myr as the profiles are equivalent to the black dotted lines. To minimize the contamination of the stripped gas from the outer region, we measure the profile at $|z|<0.5\,$kpc. After stripping, the gas column densities in $R\lesssim1\,$kpc are more enhanced compared to the control cases (black solid). The disks are truncated more by the magnetized winds (\wmhd) than by non-magnetized winds (\texttt{wHD}).}
\label{fig:disk_gas_profile}
\end{figure}

\subsection{Radial Profiles of the RPS Disk}
The different stripping of the ISM or density distribution between the two wind cases also suggests different disk truncations. In Figure~\ref{fig:disk_gas_profile}, we measure the gas column density profiles before the arrival of the winds ($t\sim100\,$Myr, black dotted) and $\sim200\,{\rm Myr}$ after the galactic disk is stripped by ram pressure ($t\sim340\,{\rm Myr}$, solid). 
Again, the black lines correspond to the profiles at the two epochs in the absence of ICM wind, which are very similar. 

Several interesting features can be observed in this plot. First, the profiles begin to diverge sharply from those in isolation at $R\sim1\,$kpc, and the break radius tends to be smaller in the MHD wind runs than in the HD wind cases. This is expected because the disk gas is stripped more by the MHD winds than by the HD winds (Figures~\ref{fig:gas_disk}--\ref{fig:disk_gas_pdf}). Second, within $R\sim1\,$kpc, the disk gas column density is rather enhanced after encountering the winds. This is consistent with \citetalias{lee20} and \citet{zhu24}, who find that the ram pressure on the RPS disks drives flows toward the galactic center, thereby increasing the fraction of dense gas in the central region. \citet{ramos-martinez18} argue that a radial inflow of gas is also produced at the interface between an ICM wind and a magnetized disk owing to an oblique shock, which gradually changes the disk gas density profile. Indeed, in MHD wind runs, where the galactic disks are strongly magnetized by the ICM, the gas column density in the center is enhanced even more. However, we find that it is difficult to separate the process proposed by \citet{ramos-martinez18} from the inflows driven by stronger disk truncation. 

In contrast, we find that the presence of the magnetic fields in the ISM itself has little impact on the ram pressure stripping. The \gbonewhd\ galaxy has slightly more gas than the \gbzerowhd\ galaxy at high densities at $t\sim 340\,{\rm Myr}$, however, this is not substantial (Figure~\ref{fig:disk_gas_pdf}). As shown in Figure~\ref{fig:pressure_rho_two} (bottom left and middle panels), the contribution of the magnetic pressure to the total pressure in \gbone\ is minor in these runs, and we do not expect disk stripping to be markedly different.


We estimate the truncation radius ($R_t$) within which the gravitational restoring force maintains its structure against the ICM pressure $P_{\rm ICM}$ using the Gunn-Gott criterion~\citep{gunn72}: 
\begin{equation}
\begin{split}
P_{\rm ICM}= & -\Sigma (R) \partial \Phi (R,z)/\partial z,
\end{split} 
\end{equation}
where $\Phi (R,z)$ is the gravitational potential measured from the total matter distribution (gas, stars, and dark matter) at a radius $R$ and vertical height $z$ in a cylindrical coordinate system, and $\Sigma (R)$ is the gas column density at $R$. {We define the ICM pressure $P_{\rm ICM}$ as the sum of ram pressure $P_{\rm ram}$, thermal pressure $P_{\rm th,ICM}$, and magnetic pressure $P_{\rm B, ICM}$, following \citet{choi22}:}
\begin{equation}
\begin{split}
P_{\rm ICM}\equiv & P_{\rm ram}+P_{\rm th,ICM}+P_{\rm B,ICM} \\  = & \rho_{\rm ICM}(v^2_{\rm ICM}+c^2_{s,{\rm ICM}})+|\mathbf{B}_{\rm ICM}|^2/\mu_0,
\end{split} 
\end{equation}
{where $\rho_{\rm ICM}$ is the ICM density, $c_{s,{\rm ICM}}$ is the sound speed of the ICM, $\mathbf{B}_{\rm ICM}$ is the magnetic field of the ICM, $\mu_0$ is the vacuum permeability.}
In this study, the magnetic pressure $P_{\rm B,ICM}$ only contributes less than $1/100$ of the total ICM pressure $P_{\rm ICM}$. We measure the galaxy column density in concentric annuli $[R,R+\Delta R]$ and a height $|z|<3\,$kpc, based on our definition of the galactic disk. We balance the gravitational restoring force against the ram pressure at a disk scale height $H$ within the half-mass radius of a gaseous disk.
At $t=100\pm25\,$Myr, the \gbzero\ and \gbone\ galaxies have mean truncated radii $R_t=$ $1.75\pm0.35\,$kpc and $1.82\pm0.30\,$kpc, respectively. These radii agree reasonably well with the break radii of the gas column densities, as shown in Figure~\ref{fig:disk_gas_profile}. The gas components at $R>R_t$ are mostly stripped within 125 Myr for both galaxies.

\subsection{Disk Star Formation Activity}
\label{sec:disk_star_formation}

\begin{figure}
\centering 
\includegraphics[width=\linewidth]{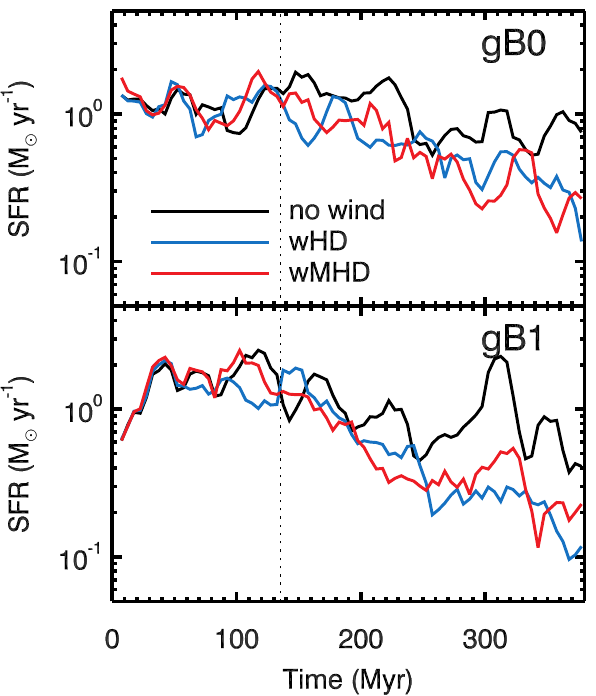}
\caption{Disk star formation rates averaged over the last 20\,Myr for the \gbzero\ (top) and \gbone\ (bottom) galaxies with no winds (black), HD winds (blue, \texttt{wHD}), and MHD winds (red, \texttt{wMHD}). The vertical dotted line indicates the time when the winds cross half the box length (150 kpc). The ICM winds effectively suppress the star formation activity. While cold gas (HI+H$_2$) is stripped more by the magnetized winds, this is not clear in the SFR evolution.}
\label{fig:sfr_disk}
\end{figure}

Strong ram pressure suppresses the disk star formation activity within a few hundred million years~\citepalias[e.g., see][]{lee20}. Figure~\ref{fig:sfr_disk} shows the star formation rates in the disks of the simulated galaxies as a function of time. After the arrival of the ICM winds, the SFRs begin to decrease and are eventually reduced by a factor of $\sim 3$ in the \gbzero\ and \gbone\ galaxies compared to those in their no-wind counterparts (black solid lines). Interestingly, the decreasing trend of the SFRs is similar for \whd\ and \wmhd, although the amount of stripped gas is greater in the \wmhd\ runs. 

\begin{figure}
\centering 
\includegraphics[width=\linewidth]{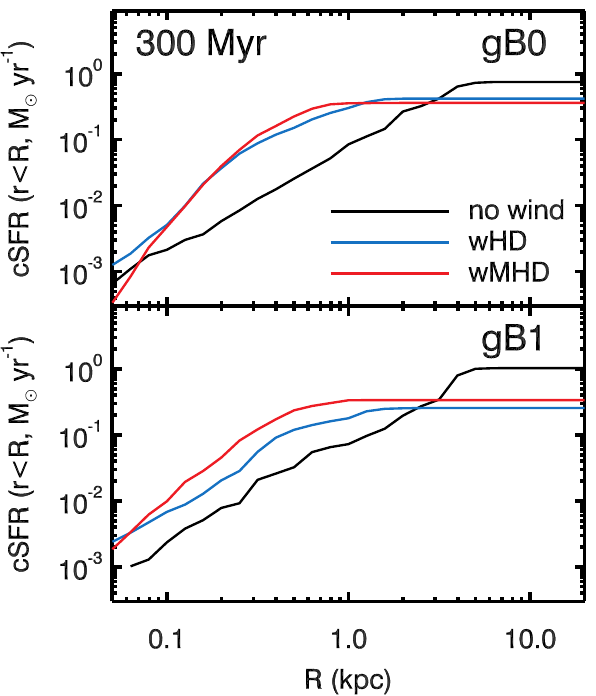}

\caption{Cumulative radial disk SFR for \gbzero\ (top) and \gbone\ (bottom) at $t=340\pm40\,$Myr. We measure the radial SFRs within $|z|<0.5\,$kpc. Both the galaxies demonstrate cumulative SFR profiles notably higher than their no-wind counterparts at their center ($R\lesssim1\,$kpc), as implied in the gas column density (Figure~\ref{fig:disk_gas_profile}). The central SFR is boosted slightly more in the \texttt{wMHD} runs. Along with Figure~\ref{fig:disk_gas_profile}, this figure reveals why the galaxies in the \texttt{wMHD} runs have SFRs similar to those in the \texttt{wHD} runs despite larger gas stripping.}
\label{fig:sfr_profile}
\end{figure}

\begin{figure*}

\centering 
\includegraphics[width=1.2\textwidth, angle=270]{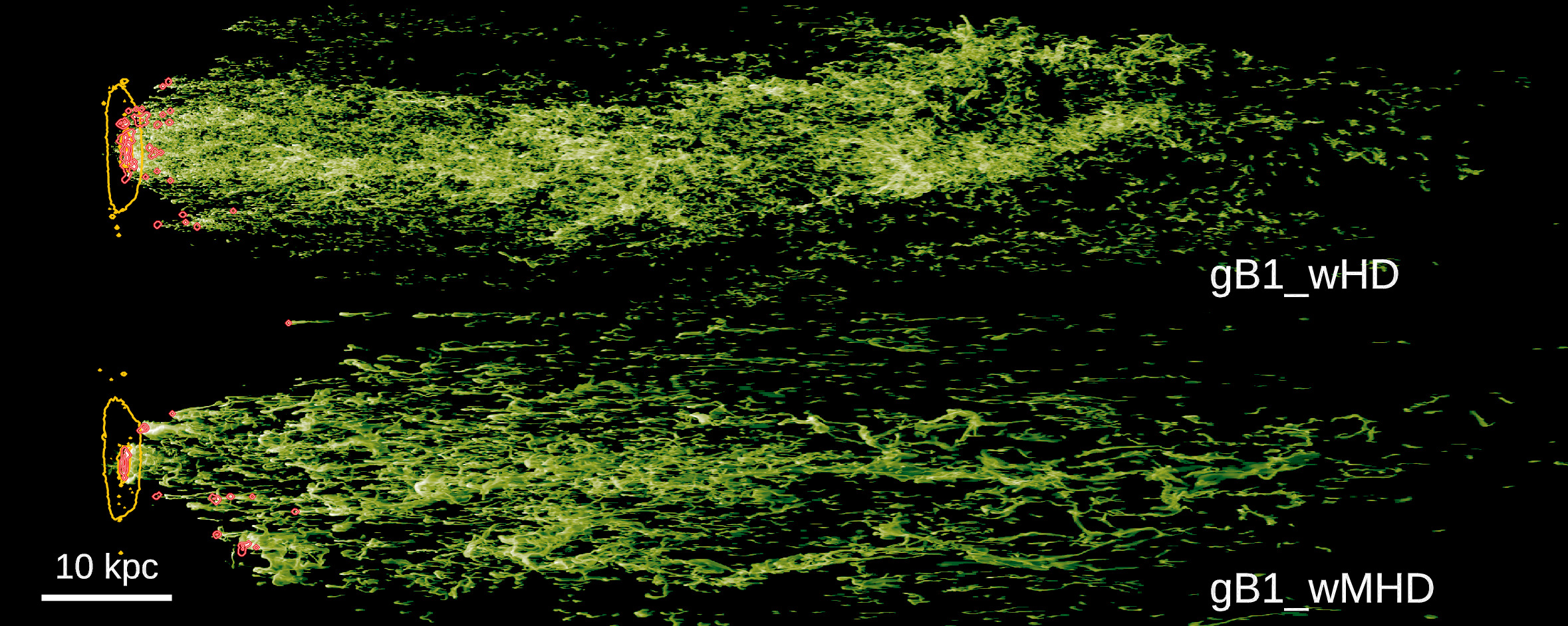}
\includegraphics[width=1.2\textwidth, angle=270]{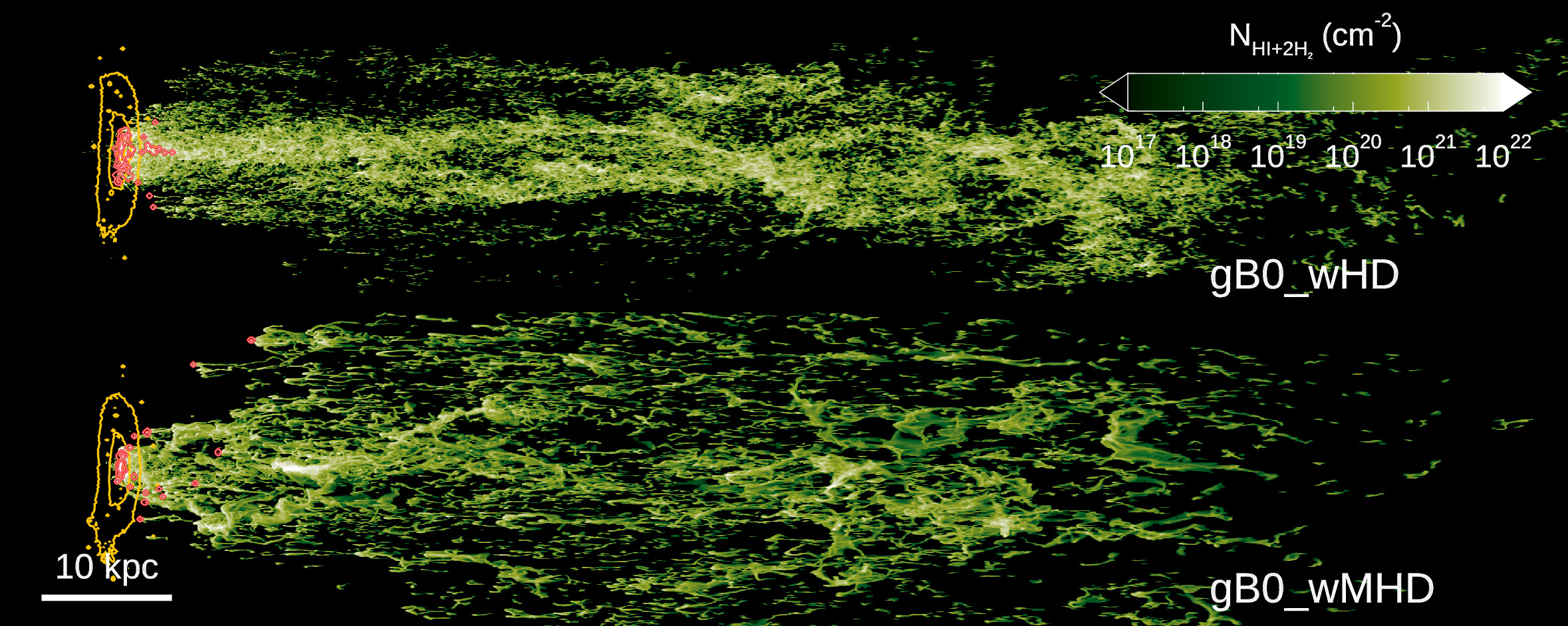}

\caption{Cold gas (HI+H$_2$) column densities of the RPS galaxies at $t=300~$Myr, $165\,$Myr after the winds start to influence the galaxies. Yellow and red contours respectively show the disks of all stars and stars younger than 20~Myr. Cold gas tails are more diffused and less fragmented in the case of magnetized winds. Clumpy clouds are efficiently formed in the distant tails when the winds are not magnetized.}
\label{fig:gas_tail}

\end{figure*}

\begin{figure}
\centering 
\includegraphics[width=0.95\linewidth]{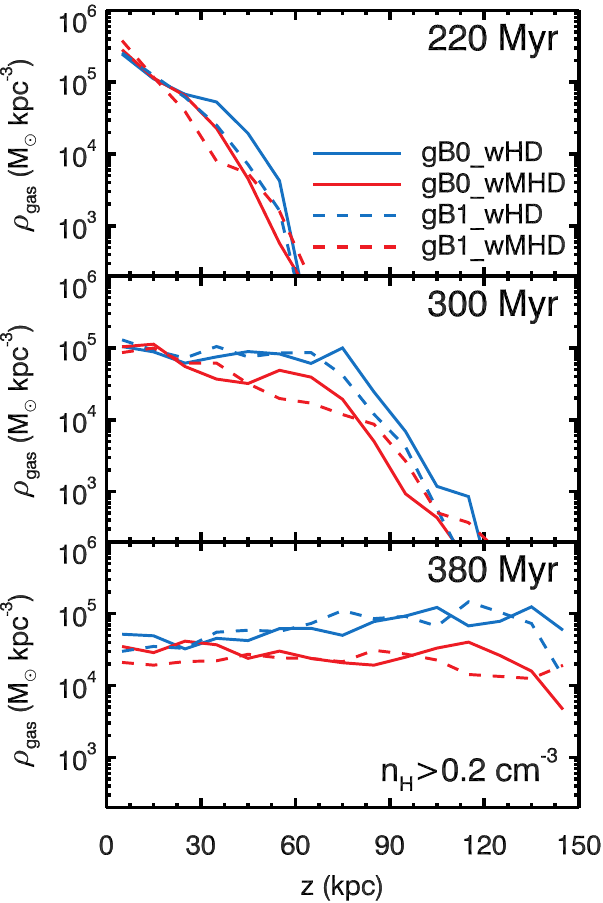}
\caption{Gas density for tail clouds with $n_{\rm H}>0.2\,{\rm cm^{-3}}$ as a function of vertical distance from the galactic mid-plane. The tail region is defined as a cylindrical volume with a radius $r=12\,$kpc and a height $3\,\text{kpc}<z<150\,$kpc from the galactic mid-plane. The density threshold corresponds to a freefall time of $t_{\rm ff}<100\,$Myr. Blue and red lines represent the gas density of the \whd\ and \wmhd\ runs, respectively, while solid and dashed lines indicate that of the \gbzero\ and \gbone\ galaxies, respectively. The amount of dense gas is similar between the \whd\ and \wmhd\ cases in the near wakes, and \wmhd\ gradually diverges from \whd\ with increasing distance and time.}
\label{fig:m_dense_tail}
\end{figure}

\begin{figure}
\centering 
\includegraphics[width=\linewidth]{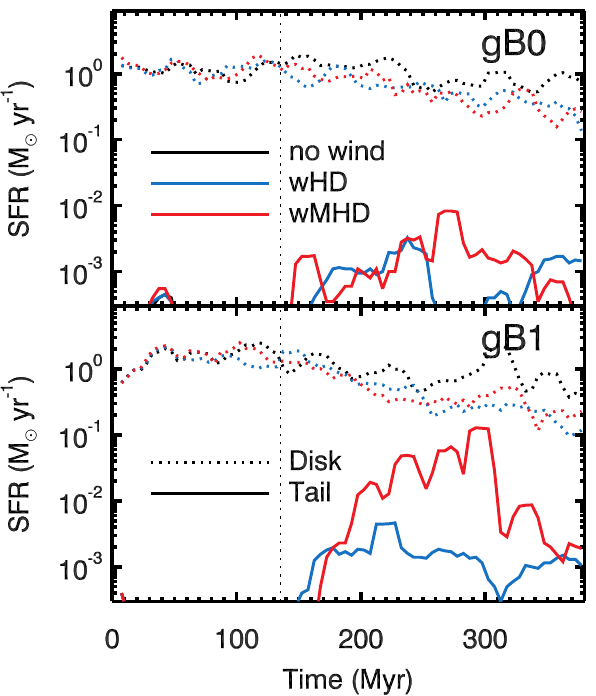}

\caption{Star formation history in the tails (solid) and disks (dotted) of the three galaxies with no winds (black), non-magnetized winds (blue, \texttt{wHD}), and MHD winds (red, \texttt{wMHD}). The top and bottom panels show the SFR for the galaxies with (\gbone) and without magnetic fields (\gbzero). The vertical dotted line indicates the time of wind arrival, as in Figure~\ref{fig:sfr_disk}. 
Galaxies exposed to the magnetized winds exhibit more star formation in their tails.}
\label{fig:sfr_all}
\end{figure}

Figure~\ref{fig:disk_gas_profile} hints at why the \whd\ and \wmhd\ runs show a similar decrease in SFRs despite the different disk gas stripping. While the outskirts of the \wmhd\ disks are truncated at smaller radii, their central gas density is increased, which can induce more star formation in the central region, as shown by \citetalias{lee20} and \citet{zhu24}. To assess this in our simulations, we compute the cumulative SFRs in the radial direction of the disk in Figure~\ref{fig:sfr_profile}. To isolate the disk star formation from off-plane star formation, we restrict our analysis to $|z|<0.5\,$kpc. When the ICM winds are imposed, star formation at $R\lesssim1\,$ kpc is significantly enhanced compared to their no-wind counterparts. This enhancement appears slightly stronger in the \wmhd\ cases, although the difference is marginal. The cumulative SFR starts to flatten out at smaller radii in the \wmhd\ runs compared to the \whd\ runs, however, the level of SFR boost is high enough to compensate for the SF truncation at the smaller stripping radii ($R\gtrsim1\,$kpc). We also confirm that the \wmhd\ galaxies have, on average, lower (local) virial parameters in the central region than those in the \whd\ runs. This condition causes more efficient star formation at $R\lesssim 1\,{\rm kpc}$, which accounts for the similar SFR evolution between the \whd\ and \wmhd\ runs in Figure~\ref{fig:sfr_disk}.


\section{Formation of RPS Tails}
\label{sec:rps_tails}

Jellyfish galaxies are characterized by unique tail features. \citetalias{lee22} demonstrate that a gas-rich disk can form a prominent RPS tail when encountering a strong ICM wind. In their simulations, a large amount of the ISM is stripped off by wind, mixing with the ICM and forming abundant, relatively dense ($\nH\gtrsim 0.1\,{\rm cm^{-3}}$), and warm ($T\sim 10^4$K) clouds. Because warm clouds can have cooling times shorter than a hundred Myr, jellyfish galaxies sometimes form clumpy molecular clouds in a distant tail before being ionized or dissipated by the hot surrounding medium. Figure~\ref{fig:gas_tail} illustrates this example. We plot the cold gas (HI+H$_2$) column density and the stellar mass distribution of the RPS galaxies at $t=300\,$Myr when the wind front reaches the boundary of the simulated box. As the simulated galaxies in this study are gas-rich, their tails are prominent in all the cases. When MHD winds are imposed, the tail clouds appear less fragmented and more spread than those in the \whd\ runs. The two cases differ in the amount of tail gas, as well as in the cloud features. Figure~\ref{fig:m_dense_tail} shows the gas density in the RPS tails for dense clouds with $n_{\rm H}>0.2\,{\rm cm^{-3}}$ (corresponding to $t_{\rm ff}<100\,$Myr). The amount of dense cloud is similar in the near wake, but gradually differs with increasing distance and time between the \whd\ and \wmhd. The two cases are therefore expected to have different star-forming features in the tail.
In this section, we first present the effect of the magnetic fields on tail star formation and then discuss the detailed stripping process in the tail.

 \begin{figure*}
\centering 
\includegraphics[width=0.85\linewidth]{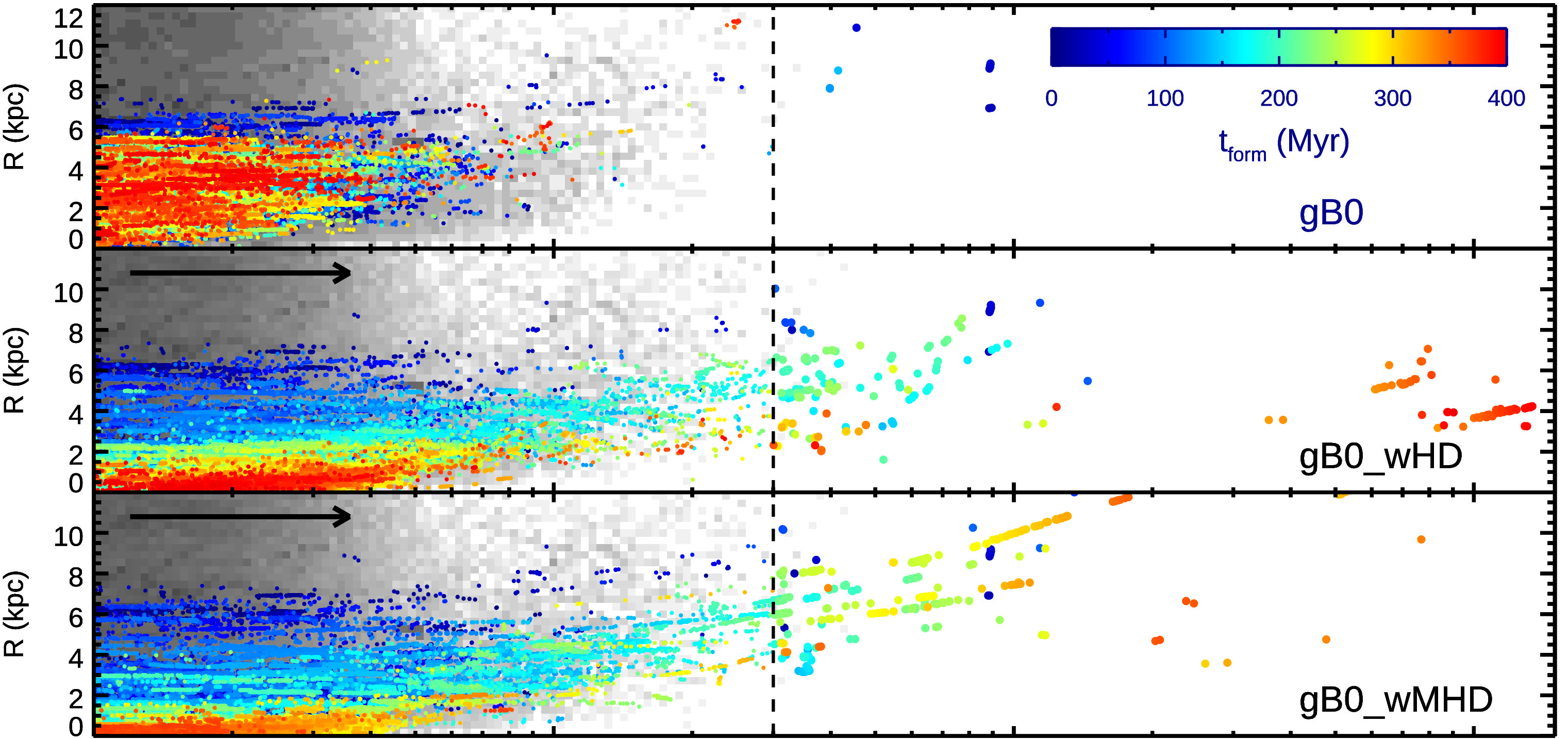}
\includegraphics[width=0.85\linewidth]{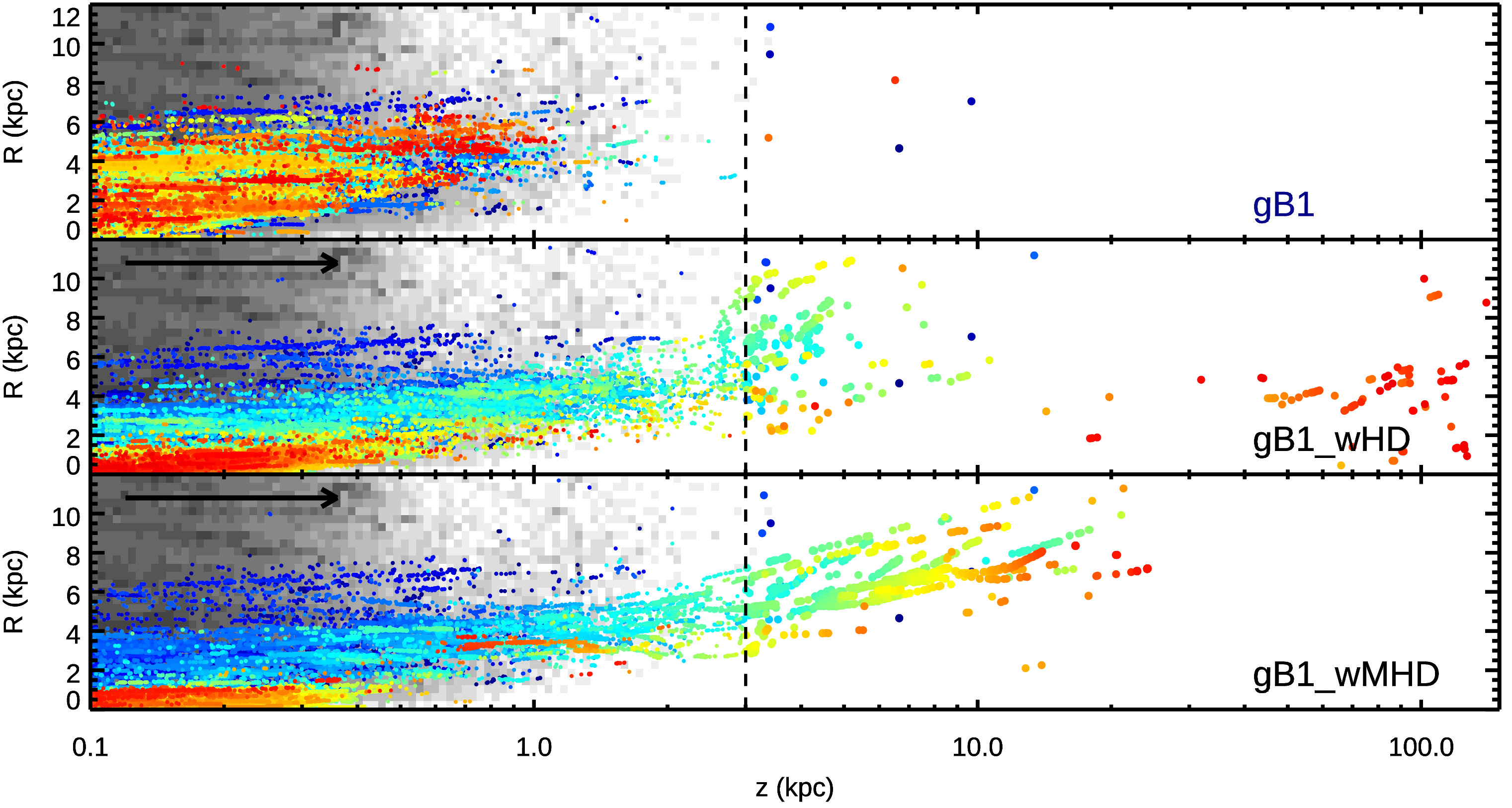}

\caption{Birthplace of stars in the cylindrical coordinates in all the simulations listed in Table~\ref{tab:ic}. The coordinate systems are defined from the center of stellar mass and the galactic mid-plane is chosen in the $XY$ plane. The black arrows denote the direction of ICM winds when included. The birth epoch of the stellar particles are displayed by color codes. The distribution of the stellar particles formed before $t=0$ is illustrated by grey shades. The vertical dashed lines mark the height of the disk ($z=3\,$kpc). Tail star formation predominantly occurs in distant tails ($z\sim50-150\,$kpc) when the winds are not magnetized. In contrast, the majority of tail stars form in the vicinity of the disk ($z\la 10\,{\rm kpc}$) during the interaction with MHD winds.}
\label{fig:sf_region}
\end{figure*}

\subsection{Star Formation in the RPS Tail}
Figure~\ref{fig:gas_tail} shows that star formation (red contours) occurs at the heads of the comet-like clumps near the tails particularly in the MHD wind cases. To quantify the tail SFR, we measure the total stellar mass formed on the disk and tail over the last 10 Myr in Figure~\ref{fig:sfr_all}. The RPS tails are measured within a cylindrical volume of radius $r<12\,$kpc and height $3<z<150\,$kpc from the galactic mid-plane.

\begin{figure*}
\centering 
\includegraphics[width=0.95\textwidth]{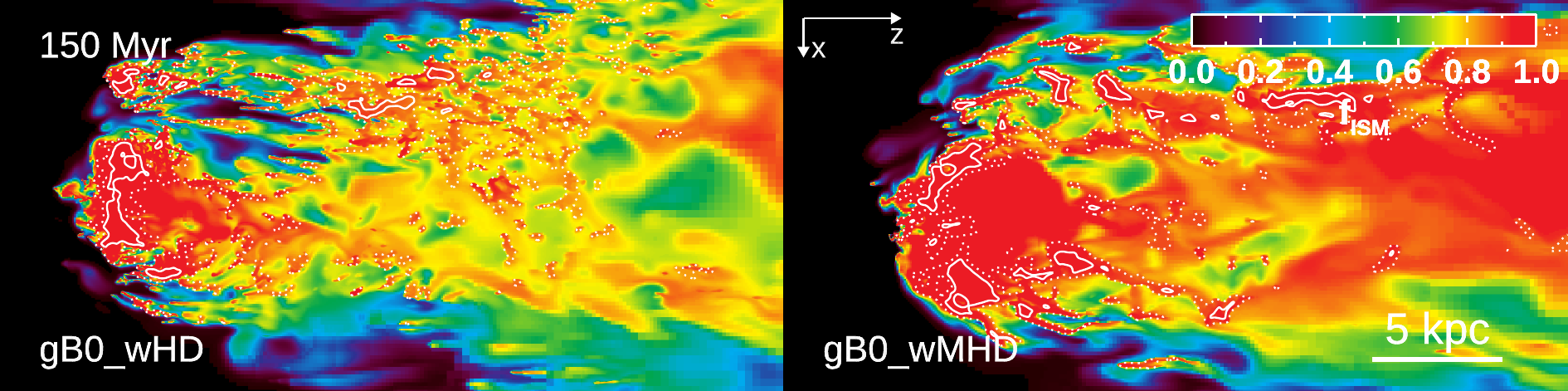}
\includegraphics[width=0.95\textwidth]{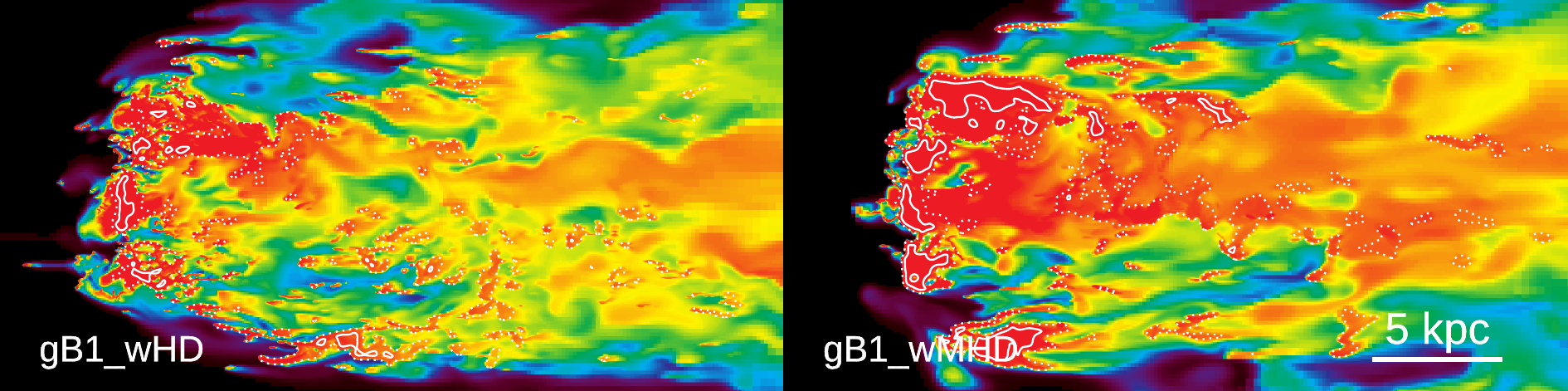}
\caption{Fraction of gas originating from the ISM in the central slab ($-250 \leq x \leq 250 \,{\rm pc}$ from the center of stellar disk) at $t=150\,$Myr for \gbzero\ (top) and \gbone\ (bottom), with HD winds (left) and MHD winds (right). The white dotted and solid contours respectively denote cold gas (HI+H$_2$) column densities of $N_{\rm H}=10^{15}$ and $10^{20}\,{\rm cm^{-2}}$. The stripped ISM does not efficiently mix with the ICM when the galaxies encounter the MHD winds.} 
\label{fig:disk_fism}
\end{figure*}

A significant level of tail SF activity is observed in all the simulations. Tail SF is particularly pronounced in the \wmhd\ runs, with higher peaks in the SFR than in the \whd\ runs. For example, the \gbonewmhd\ galaxy exhibits a peak SFR of up to 0.1 \msunyr, while in \gbonewhd\ it is an order of magnitude smaller. {This seems to contradict Figure~\ref{fig:m_dense_tail}, which shows a larger amount of dense clouds in the RPS tails of the \whd\ runs. However, the higher SFR in the \wmhd\ runs arises mostly from the near wake regions ($z<30\,$kpc), where the amount of dense clouds is comparable to or even larger than, that in the \whd\ runs.}  We partly attribute this to the fact that more ISM is stripped away in the \wmhd\ runs (Figure~\ref{fig:gas_disk}). Moreover, instabilities are suppressed by magnetic fields~\citep[e.g.,][]{frank96,ryu00}; therefore gas clouds are not efficiently fragmented and mixed with the hot ICM, but retain their cool and high-dense phase in the near wake of the \wmhd\ runs. Consequently, SF occurs in the vicinity of the galaxy within $z\lesssim 30\, {\rm kpc}$ more in the \wmhd\ cases. This is shown in Figure~\ref{fig:sf_region} where we plot the birthplaces of stars. The stream of new stars is a notable feature present only in the \wmhd\ runs. This reveals the trajectory of the star-forming clouds that are weakly mixed with the ICM due to strong magnetic fields. The SF in the near tail is also found in the \whd\ runs in the early phase of the ICM--ISM interaction. However, we find that many star particles also form in distant tails ($z\sim50$--$150\,$ kpc) at $t\gtrsim 300\,{\rm Myr}$, which is hardly observed in the magnetized ICM (Figure~\ref{fig:sf_region}). The total stellar mass formed during the late phase is not significant, and exhibits an SFR of up to $10^{-3}\,\msunyr$ (Figure~\ref{fig:sfr_all}), however, it is interesting to note that there is a gap between the stars formed in the near wake and those collapsed from the stripped gas. \citetalias{lee22} shows that mixing between the ICM and the ISM stripped from a gas-rich disk can produce a substantial amount of warm ionized gas. This gas then cools, collapses, and forms stars in distant tails within a few hundred Myr. Indeed, \citet{jones22} recently discovered young, isolated clumps of stars in the Virgo cluster, which are approximately 140~kpc away from the galaxies that are most likely the source of the gas that formed the clumps. These systems are assumed to have formed in RPS clouds, as indicated by their metallicities, which are higher than those of the dwarf galaxies with similar stellar masses. An interesting prediction from our experiments is that blue stellar systems may be formed by the interaction between the weakly magnetized ICM and gas-bearing galaxies, which deserves further investigation in the future.

\subsection{Impact of Magnetic Fields on the RPS Clouds}


The distinctive characteristics of the tail star formation, as depicted in Figure~\ref{fig:sf_region}, indicate different mixing processes. In this subsection, we investigate the differences in the properties of RPS clouds produced by the two types of wind.

\subsubsection{Mixing of Stripped ISM with ICM in the Tails}

To understand ICM--ISM mixing, we first measure the fraction of gas coming from the ISM ($f_{\rm ISM}$) by comparing the metallicity of ISM ($Z_{\rm ISM}=0.75\,Z_\odot$) and ICM ($Z_{\rm ICM}=0.3\,Z_\odot$) to the metallicity of the medium. As metal enrichment by stars is not permitted in our simulations, we define the fraction of gas originating from the ISM as $f_{\rm ISM}=(Z-Z_{\rm ICM})/(Z_{\rm ISM}-Z_{\rm ICM})$. 

\begin{figure}
\centering 
\includegraphics[width=0.95\linewidth]{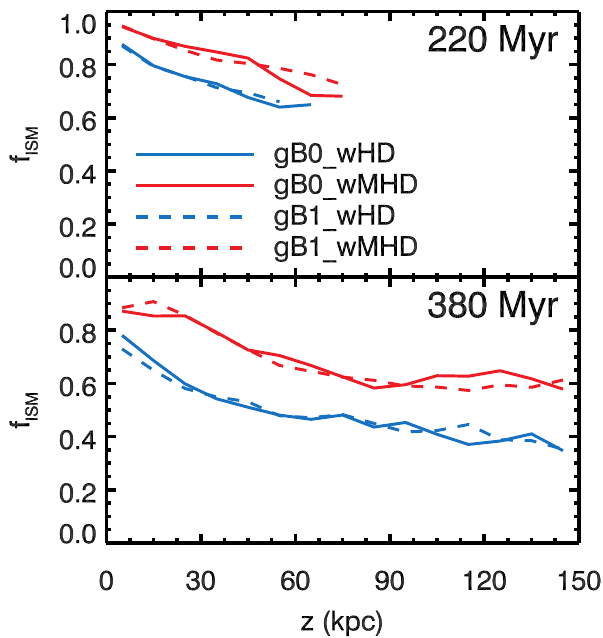}
\caption{Fraction of gas from the ISM in tail clouds with density $n_{\rm H}>0.2\,{\rm cm^{-3}}$ as a function of vertical distance from the galactic mid-plane. 
The color code and style of the lines are the same with those in Figure~\ref{fig:m_dense_tail}. The stripped ISM mixes with the ICM less efficiently when the galaxies encounter MHD winds. On the other hand, at a given wind, there is no significant difference between \gbzero\ and \gbone.}
\label{fig:fism_dist}
\end{figure}

Figure~\ref{fig:disk_fism} shows that in the early phase of the ICM--ISM interaction ($t=150\,{\rm Myr}$), $f_{\rm ISM}$ is generally very high in the RPS disk and tail clouds, often exhibiting the value close to unity in the high column density regions ($N_{\rm HI}>10^{15}\,{\rm cm^{-2}}$, marked as white dotted contours). In lower-density regions, $f_{\rm ISM}$ is reduced to 0.5--0.6, indicating that the ISM is not simply stripped and diffused, but also mixes with the ICM.

Figure~\ref{fig:disk_fism} also shows that $f_{\rm ISM}$ is generally higher in the \wmhd\ galaxies (right panels) than in the \whd\ galaxies (left panels), suggesting that mixing is less efficient. When the gaseous disk interacts with the magnetized wind, the stripped ISM is less fragmented and, as indicated by the color codes, mixes much less with the ICM. However, it is also possible that the high $f_{\rm ISM}$ could be owing to the amount of ISM entrained by the MHD winds, as shown in Figure~\ref{fig:gas_disk}. Therefore, we also examine the degree of mixing of the tail clouds as a function of vertical distance and time in Figure~\ref{fig:fism_dist}. Note that we consider only gas with densities $\nH\ge0.2\,\cmq$, which corresponds to a freefall time of $t_{\rm ff}\lesssim100\,$Myr, as such gas is directly relevant to star formation in the tail. \citetalias{lee22} demonstrate that a large amount of the ISM stripped off by strong winds mixes well with the ICM and produces numerous ionized clouds capable of cooling within $\sim 100\,{\rm Myr}$. They argue that dense ionized clouds with a cooling timescale of $t_{\rm cool}\le100\,$Myr condense into molecular clouds and eventually form new stars in the RPS tails. Figure~\ref{fig:fism_dist} further shows that the potential source of star formation in the tail is more likely to originate from the ISM when the stripped tail is strongly magnetized, whereas the ICM is the more important source for weakly magnetized wind cases owing to different degrees of mixing. 
We also note that \gbonewmhd\ forms 12 times more stars than \gbzerowmhd\ in the near wakes ($z=3-30$~kpc) after the wind arrival ($t>100$~Myr), while \gbzerowmhd\ forms only 27\% more stars than \gbzerowhd\ in the same region. However, both \gbzero\ and \gbone\ runs with the magnetized winds have a comparable amount of warm ionized gas in their tails due to similar mixing, as shown in Figures~\ref{fig:m_dense_tail} and \ref{fig:fism_dist}. This suggests that the absence of distant star formation in the magnetized wind runs is unlikely to be due to star formation in the near wakes, but is likely the result of less efficient mixing and collapse caused by magnetic fields. It is also worth noting that a significant fraction of the ISM ($f_{\rm ISM}\sim 0.4$--$0.6$) contributes to the formation of dense gas even in the distant tails ($z\gtrsim 100\,{\rm kpc}$) in both the magnetized and non-magnetized wind cases. Because the ISM is more metal-rich than the ICM, the high $f_{\rm ISM}$ also suggests that most metals in the distant tail originate from the RPS galaxy. Again, the pre-existing magnetic field in the galactic disk does not significantly alter these mixing features.

\begin{figure}
\centering 
\includegraphics[width=0.95\linewidth]{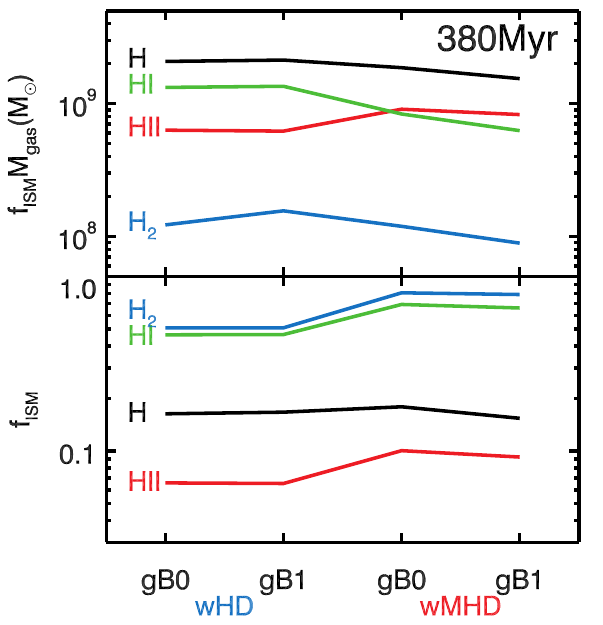}

\caption{Top: amount of ISM-origin hydrogen in the {\em tails} at the last stage of the simulations ($t=380\,$Myr). Black, red, green and blue respectively indicate the mass of H, HII, HI, and H$_2$ stripped from the ISM. In the tails of the \wmhd\ runs, the HII mass is larger, and the cold gas (HI+H$_2$) mass is lower than those in the HD wind runs. The stripped ISM is not well mixed with the ICM, however, it is more likely to be ionized in the MHD wind cases, as indicated by the higher HII mass. Bottom: Fraction of each hydrogen species originating from the ISM. All the hydrogen species in the tail exhibit a higher fraction of gas originating from the ISM with MHD wind runs than with HD winds, while the ISM-origin fraction of total hydrogen stays almost constant. This is because, as seen in the upper panel, the amount of ISM-origin cold gas decreases while HII mass increases.}
\label{fig:mgas_tail}
\end{figure}

We have shown in Figure~\ref{fig:m_dense_tail} that the amount of dense clouds is lower in the tails of the \wmhd\ runs. Since the disk gas is stripped more by the magnetized winds as shown in Figure~\ref{fig:gas_disk}, the lower number of dense clouds in the tails of the \wmhd\ runs may indicate more dissociation or ionization of stripped clouds. Figure~\ref{fig:mgas_tail} shows the mass and fraction of the ISM-origin gas in the final epoch ($t=380\,$Myr), focusing on its contribution to the different phases in the RPS tail. In the \whd\ simulations, the ISM-origin gas exists predominantly in neutral (HI), followed by ionized (HII) and molecular (H$_2$) forms. Efficient cooling owing to a fragmented ISM well-mixed with the ICM results in an HI-rich tail with mass $M_{\rm HI}\sim 10^9\,\msun$. Consequently, the HII of ISM origin is diminished, with over 90\% of the HII in the tail originating from the ICM. Conversely, in the \wmhd\ simulations, both the fraction of ISM-origin HII ($f_{\rm ISM}$) and its mass ($f_{\rm ISM} M_{\rm gas}$) are higher than those in \whd. This is because less HI mixes with ICM-origin HII, which results in a reduced cooling efficiency of the ICM. Consequently, the HI and H$_2$ tails are predominantly ($>60\%$) comprising ISM-origin gas in the presence of magnetized winds. In addition, the fraction and amount of ISM-origin HII is higher, while the amount of the ISM-origin HI and H$_2$ is lower in the magnetized ICM, indicating that the stripped ISM is more ionized or dissociated than that in the \whd\ case. This is consistent with Figure~\ref{fig:m_dense_tail}, where the magnetized winds have the RPS tails with a lesser amount of dense clouds.

\subsubsection{Alignment and Amplification of Magnetic Fields in RPS Tails}
\label{sec:alignment_amplification}

 \begin{figure}
\centering 
\includegraphics[width=\linewidth]{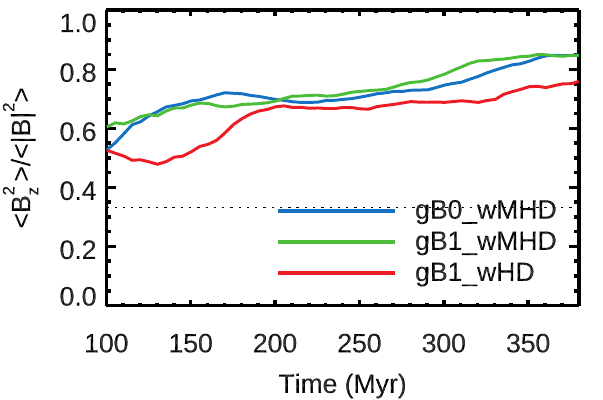}
\caption{Temporal evolution of the volume-weighted magnetic field along the $z-$axis, i.e., the wind direction, in tail gas with temperature $T<10^6\,$K for \gbzerowmhd\,(blue), \gbonewmhd\,(green), and \gbonewhd\,(red). The horizontal dotted line denotes the isotropic case. We see that the tail magnetic fields are eventually aligned with the wind.}
\label{fig:tail_mag_direction}
\end{figure}
 
Previous studies have suggested that magnetic tension stabilizes the gas flows by suppressing the growth of the Kelvin-Helmholtz instabilities~\citep[e.g.,][]{frank96,ryu00,esquivel06,heitsch09,hamlin13,liu18,praturi19}, particularly when the field lines are aligned with the flow. Figure~\ref{fig:disk_bfield} shows the alignment and amplification of the magnetic fields in a central slab with a thickness of 500~pc in the RPS tails. 
In our simulations, the initial magnetic field is perpendicular ($x-$axis) to the wind direction ($z-$axis). However, after the disks encounter MHD winds, $B_z$ components, which are aligned with the wind direction, develop in the tails. This is consistent with previous results where winds with varying magnetic fields form RPS tails with field lines well aligned with the wind direction~\citep[e.g.,][]{banda-barragan16,vijayaraghavan17,sparre20,jung23,sparre24a}. Using the GAs Stripping Phenomena in galaxies (GASP) survey with JVLA radio observations, \citet{muller21a} confirm the presence of an aligned magnetic field in the jellyfish galaxy JO206.

To quantify the magnetic field alignment in the RPS tails in the \texttt{wMHD} runs, we compute the ratio $\langle{B}^2_z\rangle/\langle{|\mathbf{B}|}^2\rangle$, where $\langle{|\mathbf{B}|}\rangle$ is the volume-weighted magnetic field strength for tail clouds ($z>3$ kpc, $R<12$ kpc) with $T<10^6\,$K, to exclude the pure ICM at $T\sim3\times10^7\,$K. Note that the isotropic magnetic fields would have a ratio of 1/3. Figure~\ref{fig:tail_mag_direction} illustrates the temporal evolution of the volume-weighted magnetic field $\langle |\mathbf{B}|^2\rangle$ of the tail gas. We find that the RPS tails have $\langle{B}^2_z\rangle/\langle{|\mathbf{B}|}^2\rangle\sim0.7-0.9$ after the tails are sufficiently developed ($t>200\,$Myr), indicating a strong alignment of the magnetic field with the wind direction, regardless of the presence of the magnetic fields in the ISM. Interestingly, the alignment is weaker in the \gbonewhd\ galaxy, suggesting that the mixing process is more turbulent than that in the \wmhd\ cases. The reorientation (and amplification) of the magnetic fields occurs immediately after the ISM--ICM interaction, with a tendency for distant tails to be more aligned with the wind direction, which is also observed in the simulated galaxies of \citet{sparre24a}. This alignment remains relatively constant over time at a given location within the tail.





Using MHD simulations of cloud--wind interactions, \citet{sparre20} argue that wind draping enhances magnetic fields, which are subsequently amplified through adiabatic compression and/or shear \citep[e.g.,][]{vainshtein72,schekochihin02,donnert18}. Similarly, we find that the disk magnetic fields are rapidly amplified from a few $\mu G$ to  several tens of $\mu G$ at $\nH\sim1\,\cmq$ in the presence of magnetized winds. Once the ISM gas is stripped, it forms the tail gas with similar $|\mathbf{B}|$ initially. However, unlike the ISM gas exhibiting the density dependence $|\mathbf{B}|\propto\rho^{2/3}$, the tails of the \texttt{wMHD} runs exhibit a weak correlation as $|\mathbf{B}|\propto\rho^{0-0.2}$. The slope is even shallower than that expected for the two-dimensional adiabatic compression ($|\mathbf{B}|\propto\rho^{0.5}$), suggesting that shear amplification is likely to play a role in the tail magnetic field amplification. 

\section{Discussion}
\label{sec:discussion}
\subsection{Effect of Magnetic Draping Layers in RPS Galaxies}
\label{sec:51}

Magnetic draping occurs when a cloud or gaseous disk moves through a magnetized medium, forming a magnetic layer at the leading edge~\citep[e.g.,][]{dursi08,pfrommer10}. This layer can inhibit the growth of instabilities, potentially reducing the gas stripping rate of the disk compared to purely hydrodynamic cases \citep{ruszkowski14,tonnesen14,ramos-martinez18}. In our \wmhd\ simulations, amplified magnetic fields are observed in the draping layers on the shock front (see the second and third rows of Figure~\ref{fig:disk_bfield}). However, we find that disk stripping is not significantly suppressed when the ISM is magnetized by wind, as shown in Figure~\ref{fig:gas_disk}. We attribute the discrepancy between our results and those of the previous studies to differences in the simulation settings and methodologies. First, in previous studies, the ram pressure is assumed to be weaker ($P_{\rm ram}/k_{\rm B}=6.4\times10^4{\rm cm^{-3}\,K}$ in \citealt{tonnesen14}, $P_{\rm ram}/k_{\rm B}=1.6\times10^5{\rm cm^{-3}\,K}$ in \citealt{ruszkowski14}) than  in our simulations ($P_{\rm ram}/k_{\rm B}=5\times10^5{\rm cm^{-3}\,K}$). To ensure a more direct comparison, we also conduct additional simulations for the \gbzero\ galaxy with both magnetized ($B_x=1\,\mg$) and non-magnetized mild winds, reducing the ram pressure to 1/10 of the fiducial $P_{\rm ram}$ by decreasing the ICM density. Although not presented here, the results confirm the formation of magnetic draping layers, with the MHD wind still stripping the disk gas more efficiently than its non-magnetized counterpart. This is also observed when mild winds are applied to the \gbone\ galaxy model. 

This leads us to consider the possibility that instabilities are suppressed; however, this effect is masked by efficient stripping owing to turbulent structures. Previous studies have not considered star formation and stellar feedback which are known to induce turbulent motions within the disk. To verify these results, we perform additional simulations using a non-magnetized disk, excluding gas cooling and star formation. Because the stellar feedback does not deplete or expel the disk gas, we assume a disk mass of $M_{\rm gas}=3.93\times10^9\,\msun$, which is consistent with the relaxed state of the disk gas in \gbzero. Again, we impose both MHD and HD winds with mild ($P_{\rm ram}/k_{\rm B}=5\times10^{4}\,{\rm cm^{-3}\,K}$) and strong ($P_{\rm ram}/k_{\rm B}=5\times10^5\,{\rm cm^{-3}\,K}$) ram pressure on the disk.

\begin{figure}
\centering 
\includegraphics[width=\linewidth]{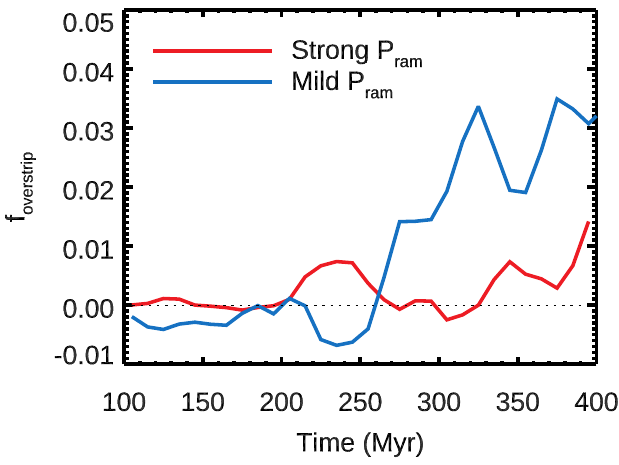}
\caption{Overstripping ratio of the disk gas removed by HD winds relative to the disk gas stripped by MHD winds for galaxies simulated with no cooling nor star formation. The ratio is normalized by total stripped mass in the MHD wind runs during $t=100-400\,$Myr. The red and blue colours denote simulations with strong and mild (i.e., 1/10 of the strong case) ram pressure. The overstripping ratio becomes larger than zero approximately $100\,$Myr after the wind arrival, indicating more gas removal by the HD winds.}
\label{fig:overstip}
\end{figure}

Figure~\ref{fig:overstip} shows the over-stripping ratio of the gas in the HD wind runs to the total stripped mass by the MHD winds during $t=100-400\,$Myr, $f_{\rm overstrip}(t)=(M_{\rm gas}^{\whd} (t)-M_{\rm gas}^{\wmhd}(t))/(M_{\rm gas}^{\wmhd} (t_f)-M_{\rm gas}(t_0))$, where $M_{\rm gas}^{\whd} (t)$ and $M_{\rm gas}^{\wmhd} (t)$ are the disk gas mass at time $t$ in the HD and MHD wind runs, respectively; $M_{\rm gas}^{\rm \texttt{wMHD}} (t_f)$ is the disk gas mass at $t_f=400\,$Myr in the MHD wind runs; and $M_{\rm gas} (t_0)$ is the disk gas mass at $t_0=100\,$Myr, right before the wind arrival. Overall, the disks lose more gas when encountering HD winds, and this trend is more evident under mild ram pressure. These results are consistent with previous studies and show that magnetic draping reduces gas stripping only when the winds are mild and the ISM unperturbed by SN explosions. Even in this case, however, the effect of the magnetic draping layer is still minor in the stripping process, resulting in at most a few percent reduction in our setup. Further details of this investigation are provided in \ref{sec:draping}.

\subsection{Different Effects of ICM winds on Smooth and Turbulent Galaxies}

In Section~\ref{sec:disk_stripping}, we demonstrated that MHD winds strip a larger gas mass of the ISM than HD winds. This is attributed to the strong magnetic pressure in the disks of the \wmhd\ runs, which alters the density distribution of the disk clouds, making the clouds more vulnerable to external perturbations (Figures~\ref{fig:disk_gas_pdf}--\ref{fig:pressure_rho_two}). However, we note that the effect of the MHD winds on gas stripping depends on the structure of the ISM.

Figures~\ref{fig:disk_projection_edgeon_ncns} and \ref{fig:disk_projection_faceon_ncns} show that the RPS features of smooth galaxies (without cooling, star formation, and stellar feedback) are hardly affected by wind magnetization. This is in stark contrast to the turbulent galaxies shown in Figure~\ref{fig:disk_projection}, where the presence of wind magnetic fields leaves distinct features in the disk and tail clouds. This difference arises from the stable ISM in smooth galaxies, supported by thermal pressure, which inhibits the permeation of magnetic fields through MHD advection because of the absence of turbulent pressure. Moreover, their disk scale-heights are approximately half those of \gbzero, resulting in a denser gaseous disk in the midplane. Therefore, the disks have slower Alfvén velocities than those in turbulent galaxies, which may further delay the magnetization of the disks.

In contrast, in active star-forming galaxies, disk clouds are often shattered and highly perturbed by stellar feedback, particularly in the outer regions, as illustrated in the left column of Figure~\ref{fig:disk_projection}. This vigorous feedback leads to rapid magnetization of clouds and the entire ISM through MHD winds. In diffuse clouds with intermediate densities, strong magnetic pressure can effectively interrupt gravitational collapse and assist ram pressure in cloud stripping. Consequently, the impact of magnetic fields on the structure of star-forming galaxies can differ substantially from that observed in smooth galaxies without cooling and star formation. 

\section{Conclusions}
\label{sec:conclusions}

We investigated the impact of magnetic fields on the formation of jellyfish galaxies using a set of idealized simulations of a dwarf galaxy by explicitly modeling the multiphase ISM through star formation and feedback. We found that the presence of magnetic fields in the ICM strongly affects RPS features. Our results can be summarized as follows.

\begin {enumerate}

\item Magnetic draping by magnetized ICM winds amplifies magnetic fields by an order of magnitude. The initial magnetic fields of the ICM wind in \wmhd\ are perpendicular to the wind direction ($v_{\rm ICM}$ in the $z$ direction while $\mathbf{B}$ aligned with the $x$ direction); however, with time, the fields in the RPS tails become predominantly aligned with the wind direction. Alignment is also observed in \gbonewhd, however, with a much weaker field strength (Figure~\ref{fig:tail_mag_direction}). The field alignment is consistent with recent radio observations~\citep{muller21b}.

\item Strong ram pressure ($P_{\rm ram}/k_{\rm B}=5\times10^5\,{\rm cm^{-3}\,K}$) removes $\sim90\%$ of the disk gas in all the simulated galaxies within 250\,Myr. Disk gas stripping is further facilitated by the MHD winds, which strongly magnetize the disks, forming magnetic pressures comparable to turbulent pressures. Magnetic fields induce force fields that act against gravity both in the stripped disks and tails (Figure~\ref{fig:f_rho}). In this case, the disk clouds become less clumpy and thus more susceptible to ram pressure. 

\item RPS clouds show smoother and less clumpy structures when embedded in magnetized ICM winds compared to pure HD winds. Star formation is preferentially localized near wakes ($z<20\,$kpc) in the \wmhd\ runs, leaving clear trails of star-forming regions (Figure~\ref{fig:sf_region}). The stripped ISM is dragged through the RPS tails with minimal mixing with the ambient ICM. The limited mixing allows the stripped ISM to sustain star formation in the near wakes ($\sim10\,$kpc) for $\sim200$\,Myr. However, the smoother morphology of the tails in the \wmhd\ runs results in the gradual dissipation of clouds at larger distances, suppressing star formation in the distant tails.

\item Simulations with HD winds show weaker star formation in the near wakes, whereas new clumps of stars form in the distant tails ($z\sim100\,$kpc). The ISM stripped by the HD winds effectively mixes with the ICM, producing numerous warm clouds in the tail. These clouds can cool and collapse within a few hundred Myr, leading to star formation sites farther from the galaxy than those with magnetized winds (Figure~\ref{fig:sf_region}).

\item  We reproduce previous findings that less efficient stripping occurs in simulations with magnetized winds but only when cooling and star formation are neglected. As noted above, this is the opposite in turbulent galaxies, highlighting the critical role of star formation and stellar feedback in studying the impact of magnetic fields on galaxies.

\item The gas density increases at the center in the RPS disks, which is more apparent in the case of the MHD winds. The density increase at the center is consistent with the result of \citet{zhu24} who find gas inflow driven by ram pressure on the disk planes. However, the magnetic fields initially imposed in the ISM ($\sim1-10\,\mg$ after relaxation, see Figure~\ref{fig:bmag_rho}) do not significantly affect gas stripping (\gbzero\ vs \gbone), likely because of their minor contribution to the internal energy budget (Figure~\ref{fig:pressure_rho_two}).

\item Strong ram pressure reduces star formation by a factor of four compared with a galaxy without wind (Figure~\ref{fig:sfr_disk}). Although MHD winds truncate gaseous disks more effectively, the decline in SFRs exhibits a similar trend in both the \whd\ and \wmhd\ runs. This similarity arises from a stronger central SF boost in the \wmhd\ runs, driven by the enhanced central gas density (Figure~\ref{fig:sfr_profile}).

\end {enumerate}

Jellyfish galaxies exhibit distinct cloud morphologies and star formation sites in their tails when interacting with the ICM through ram pressure. Given the presence of magnetic fields in the ICM, as reported by radio observations \citep[e.g.,][]{carilli02,govoni04,vogt05,bonafede10,kuchar11,bohringer16,govoni17,osinga22,Xu22}, our results suggest that magnetic fields play a key role in the formation process of jellyfish galaxies in clusters, along with factors such as the strength of ram pressure (\citetalias{lee20}) or disk gas content (\citetalias{lee22}). 

Although progress has been made, significant challenges remain unresolved. As noted by \citet{tonnesen19}, varying the ram pressure alters the stripped features because the gas density profile of the disk is gradually modified by the initial pressure. Furthermore, the stripping timescale of disk gas increases significantly when the ram pressure varies along the orbital motion in cosmological simulations~\citep{rohr23}. This underscores the need for follow-up studies that incorporate varying ram pressures and magnetic fields over the orbital timescales of cluster satellites to develop theoretical frameworks that are directly comparable to observations. Furthermore, this study does not consider thermal conduction, which has been proposed to suppress the development of hydrodynamic instabilities between two different media~\citep{vieser07a,vieser07b,armillotta16,armillotta17,bruggen16}. However, \citet{kooij21} suggest that magnetic fields guide the flow of electrons, typically aligning with the interface between two different media, which eventually lowers thermal conduction efficiency. However, as demonstrated in this study, more realistic setups can yield results that differ significantly from ideal scenarios. Therefore, future studies should incorporate the realistic physical processes of galaxies and their environments to better understand these complex interactions.

Finally, observations of of D100 and ESO~137-001 revealed intriguing HI-deficient but H$_2$ rich tails \citep{jachym14,jachym17,jachym19,ramatsoku25} that require theoretical interpretation. However, our simulations do not reproduce these high molecular to HI mass fractions. Further studies with larger parameter ranges and more realistic physical models are required to fully address this issue.

\section*{acknowledgments}
J.L. is supported by the National Research Foundation (NRF) of Korea grant funded by the Korea government (MSIT, RS-2021-NR061998). J.L. also acknowledges the support of the NRF of Korea grant funded by the Korea government (MSIT, RS-2022-NR068800). T.K. is supported by the NRF of Korea (RS-2022-NR070872 and RS-2025-00516961), and acted as the corresponding author. This work was also partially supported by the Yonsei Fellowship, funded by Lee Youn Jae. S.M.A. is supported by a Kavli Institute for Particle Astrophysics and Cosmology (KIPAC) Fellowship, and by the NASA/DLR Stratospheric Observatory for Infrared Astronomy (SOFIA) under the 08\_0012 Program. J.R. was supported by the KASI-Yonsei Postdoctoral Fellowship and was supported by the Korea Astronomy and Space Science Institute under the R\&D program (Project No. 2025-1-831-02), supervised by the Korea AeroSpace Administration. This work is also supported by the Center for Advanced Computation at Korea Institute for Advanced Study. 
This work used the DiRAC Data Intensive service (DIaL2 / DIaL [*]) at the University of Leicester, managed by the University of Leicester Research Computing Service on behalf of the STFC DiRAC HPC Facility (www.dirac.ac.uk). The DiRAC service at Leicester was funded by BEIS, UKRI and STFC capital funding and STFC operations grants. DiRAC is part of the UKRI Digital Research Infrastructure.

\appendix
\label{sec:appendix}
\restartappendixnumbering
\section{Relation between star formation rate and pressure}
\label{sec:SFR_pressure}

\begin{figure}
\centering 
\includegraphics[width=1\linewidth]{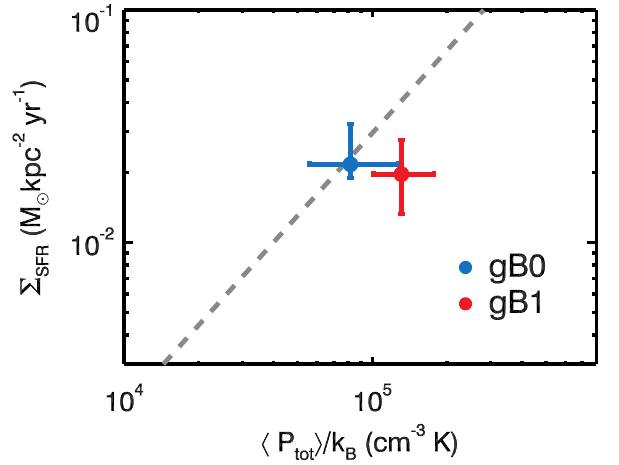}

\caption{SFR column density as a function of volume-weighted total pressure $\langle P_{\rm tot} \rangle$ in the galactic midplanes of our simulations with no ram pressure (\gbzero\ and \gbone). The total pressure is measured within the scale heights and half stellar mass radii of the disks. The scatter bars indicate the $16^{\rm th}-84^{\rm th}$ percentile scatter of the relation within $t=0-380\,$Myr. The grey dashed line shows the result from \citet{ostriker22}. The magnetized galaxy (i.e.,\gbone) has slightly higher total pressure at a given SFR column density.
}
\label{fig:sfr_pressure}
\end{figure}

\citet{kim13} and \citet{ostriker22} show a close relation between the total pressure and star formation rates on disks owing to profound effect of feedback processes on the ISM. \citet{ostriker22} suggest the fitting formula that relates volume-weighted total pressure $\langle{P}_{\rm tot}\rangle$ of cold and warm clouds ($T<2\times10^4\,K$) with the SFR column density:
\begin{equation}
\label{eqn:ostriker22}
\Sigma_{\rm SFR}=1.95\times10^{-3}\,\msun\,{\rm kpc^{-2}\,yr^{-1} } \left( \frac{\langle{P}_{\rm tot}\rangle/k_{\rm B}}{10^4\,{\rm cm^{-3}\,K}} \right)^{1.18}.
\end{equation}
Motivated by the results of \citet{ostriker22}, we examine the total pressure-SFR column density relation in our simulations with no ram pressure. The volume-weighted total pressure $\bar{P}_{\rm tot}$ is the sum of the thermal pressure, vertical Maxwell stress, and vertical turbulent pressure. The pressure is measured within the half-stellar mass radius and disk scale height $H=\sqrt{\int\rho z^2{\rm d}V/\int\rho {\rm d}V}$, where $\rho$ is the total matter density, $z$ is the vertical distance from the galactic midplane, and $dV$ is the volume of a cloud. We compute the three pressure components as follows:
\begin{equation}
\label{eqn:thermal_pressure}
\langle{P}_{\rm th}\rangle=\frac{  \int P_{\rm th}\Theta(T<2\times10^4\,{\rm K}) dV } {\int \Theta(T<2\times10^4\,{\rm K}) dV },
\end{equation}
\begin{equation}
\label{eqn:mag_pressure}
\langle{\Pi}_{\rm B}\rangle=\frac{  \int \Pi_{\rm B}\Theta(T<2\times10^4\,{\rm K}) dV } {\int \Theta(T<2\times10^4\,{\rm K}) dV },
\end{equation}
\begin{equation}
\label{eqn:turb_pressure}
\langle{P}_{{\rm turb},z}\rangle=\frac{  \int \rho v_{z}^2 \Theta(T<2\times10^4\,{\rm K}) dV } {\int \Theta(T<2\times10^4\,{\rm K}) dV },
\end{equation}
where $\Theta(X)$ is the Heaviside step function that returns 1 when the argument $X$ is true, $T<2\times10^4\,$K is the temperature threshold of cold and warm gas, $\Pi_{\rm B}\equiv(|\mathbf{B}|^2-2B_z^2)/8\pi$ is the vertial Maxwell stress, 
$\rho$ is the density of a cloud, and $v_z$ is the vertical velocity of the cloud. Figure~\ref{fig:sfr_pressure} shows the SFR column density as a function of the volume-weighted total pressure $\langle{P}_{\rm tot}\rangle$ measured for clouds with $T<2\times10^4\,K$ within a half-stellar mass radius with no wind. Following \citet{ostriker22}, the SFR column density is averaged over 40\,Myr. The gray dashed line denotes the fit given in Equation~\ref{eqn:ostriker22}. The scatter bars in Figure~\ref{fig:sfr_pressure} show the $16^{\rm th}-84^{\rm th}$ percentile distributions of the total pressure and SFR column density at $t=0-380\,$Myr. At a fixed total pressure, the SFR column density decreases with an increase in the disk magnetic field strength. 
While \gbzero\ reasonably agrees with the prediction of \citet{ostriker22}, \gbone\ exhibits a higher total pressure for a given SFR, as star formation becomes slightly more bursty~\citep[e.g.,][]{martin-alvarez25} and both turbulent and thermal pressure increase. Nevertheless, turbulent pressure dominates over thermal pressure in both galaxies  ($P_{\rm turb}/P_{\rm th}\sim 20$), allowing us to study how turbulence-supported galaxies respond to the strong, magnetized ICM winds.

\section{Magnetic draping and disk gas stripping}
\label{sec:draping}

\begin{figure}
     
\centering 
\includegraphics[width=0.95\linewidth]{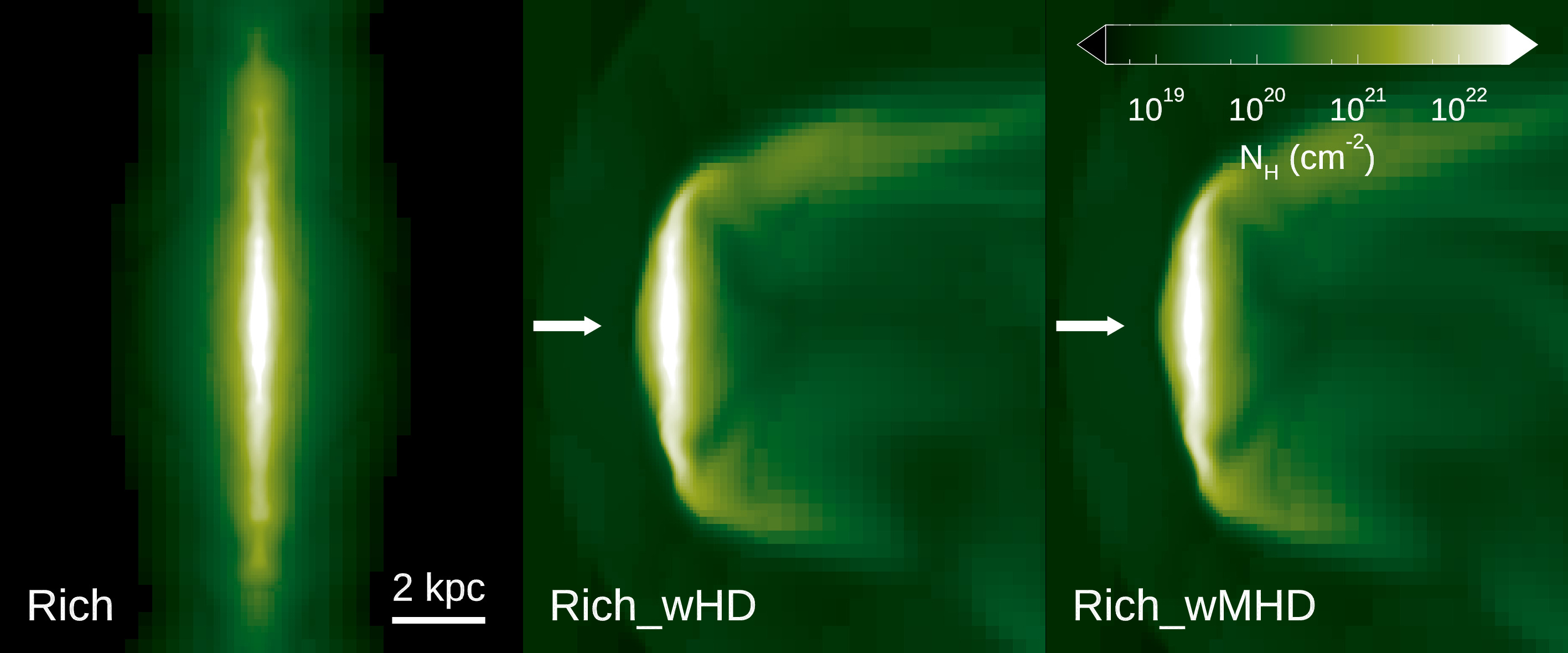}
\includegraphics[width=0.95\linewidth]{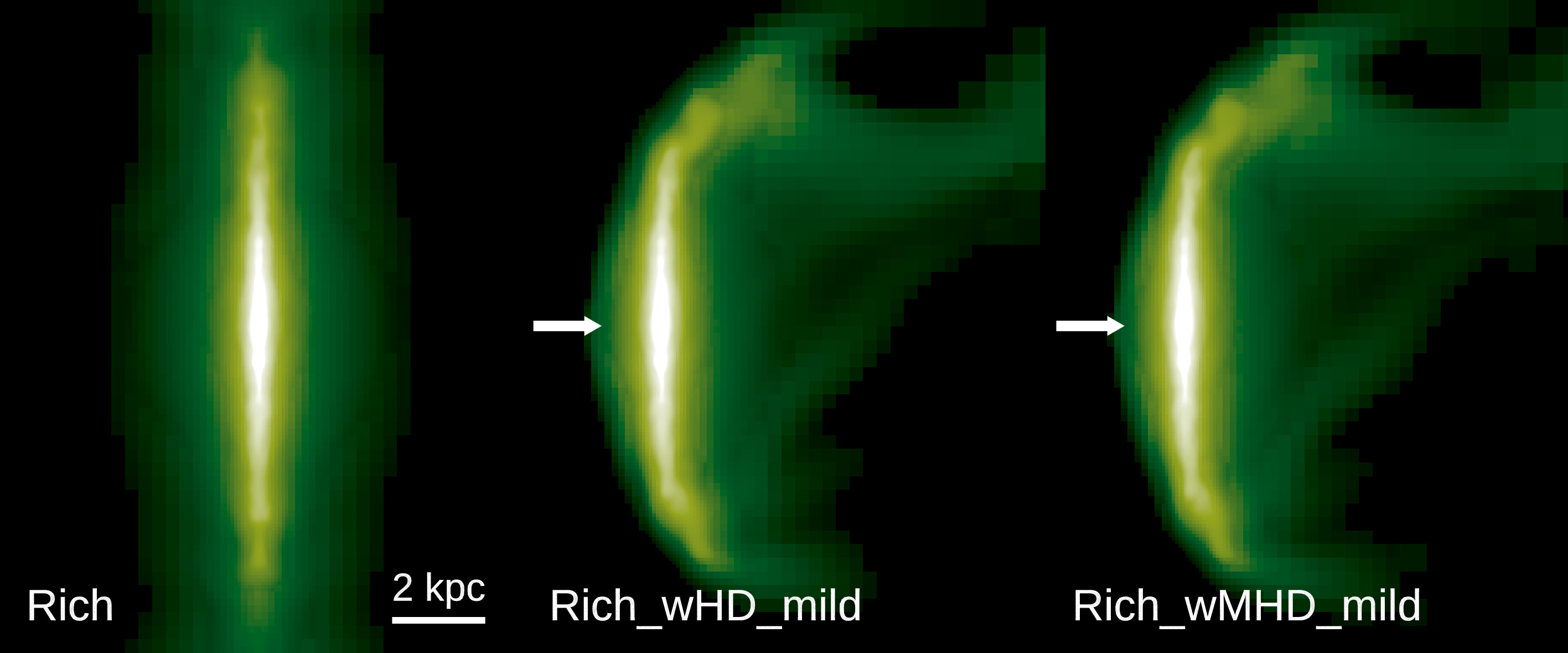}
\includegraphics[width=0.95\linewidth]{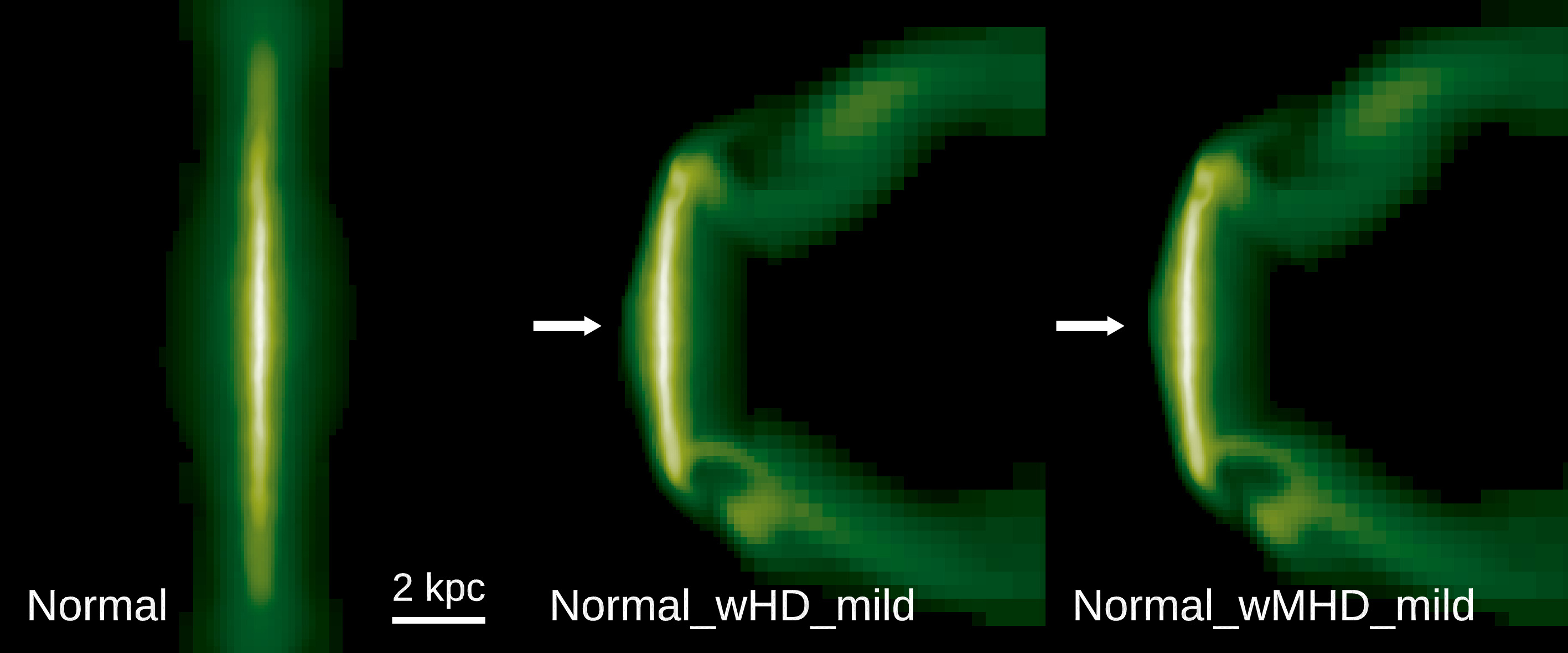}
\caption{Edge-on hydrogen column density maps of the static disks with no gas cooling and no star formation within a $\Delta x=1$kpc slab centered on the  stellar disk at $t=150~$Myr. We show the maps right after the disks encounter the strong (top row) and mild winds (1/10 of the strong case, middle and bottom rows). The top and middle rows correspond to gas-rich disks while the bottom row corresponds to the disk with a normal gas fraction. The second and third columns respectively depict simulations with non-magnetized and magnetized ($B_x=1\,\mg$) winds. White arrows indicate the direction of the winds. While tail morphology is significantly different between \whd\ and \wmhd\ simulations in Figure~\ref{fig:disk_projection}, such difference is hardly seen here.}
\label{fig:disk_projection_edgeon_ncns}
\end{figure}

\begin{figure}
\centering 
\includegraphics[width=0.95\linewidth]{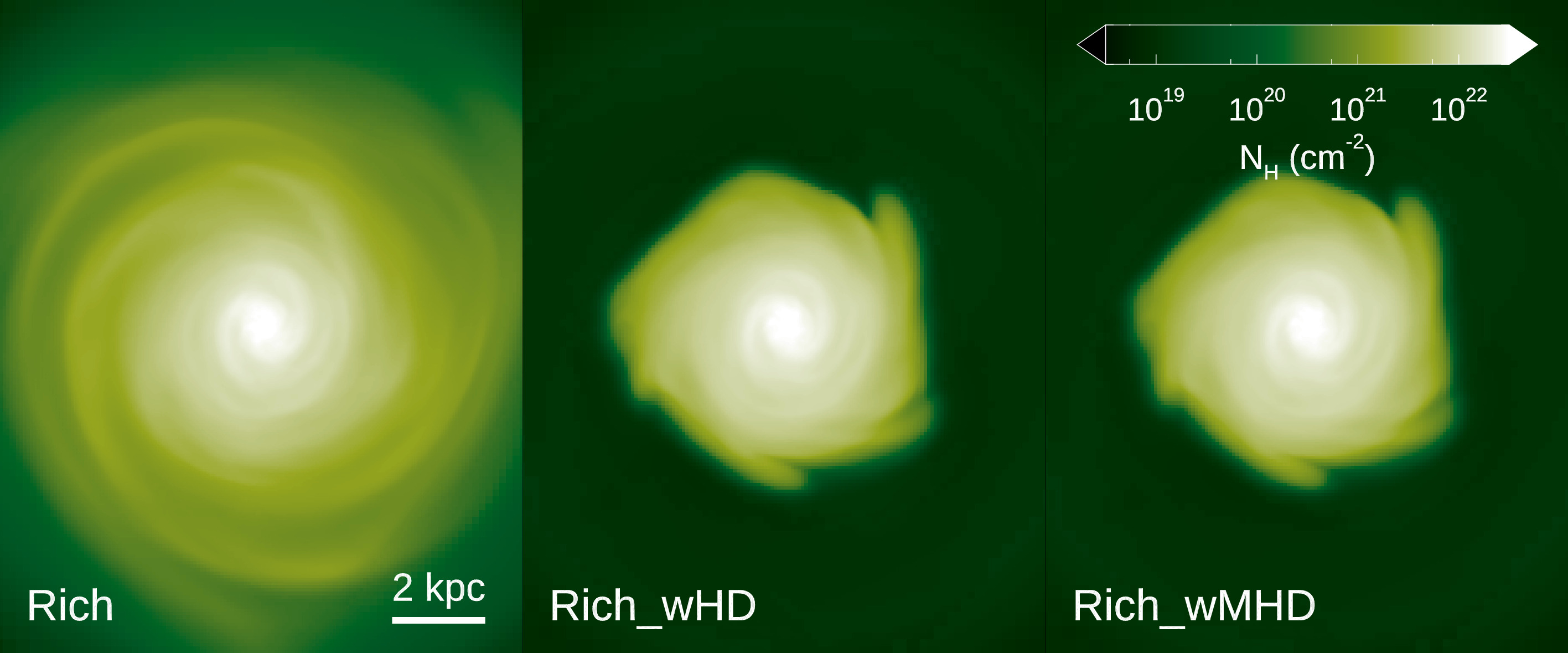}
\includegraphics[width=0.95\linewidth]{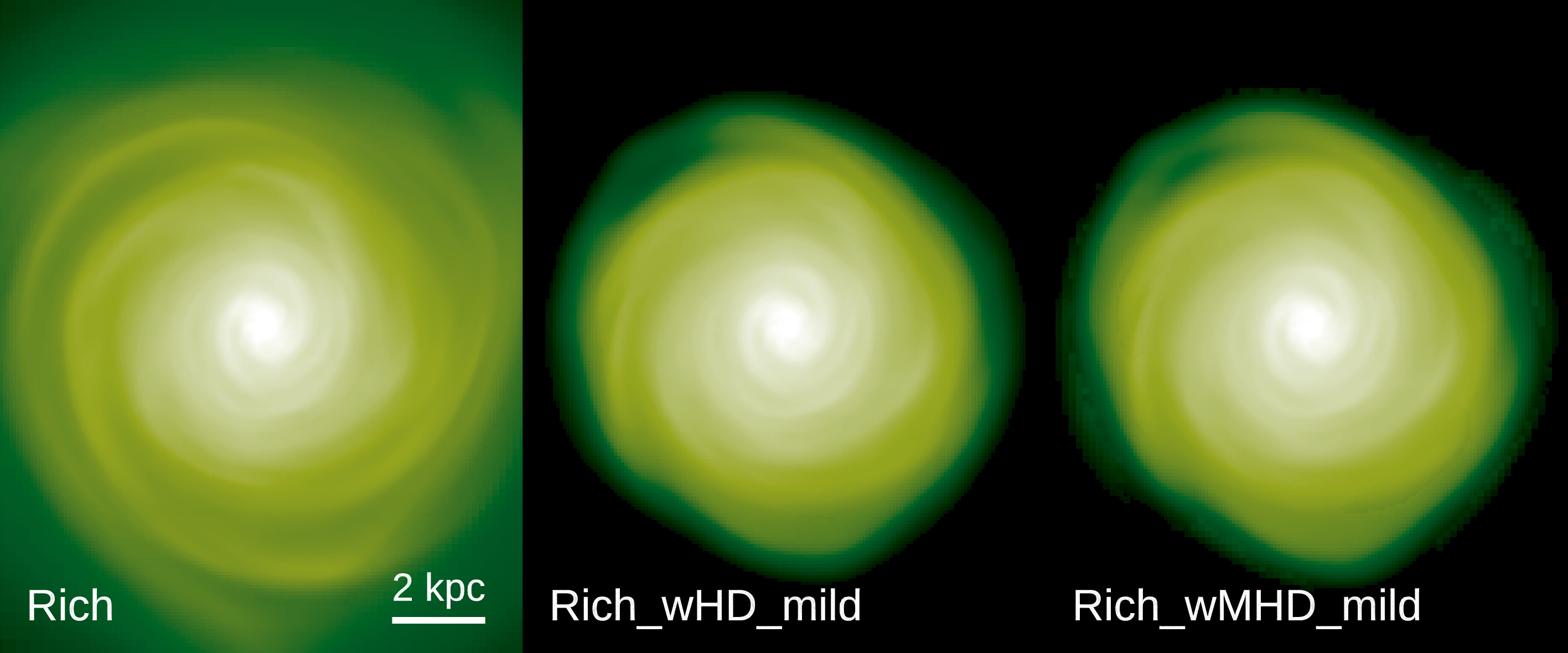}
\includegraphics[width=0.95\linewidth]{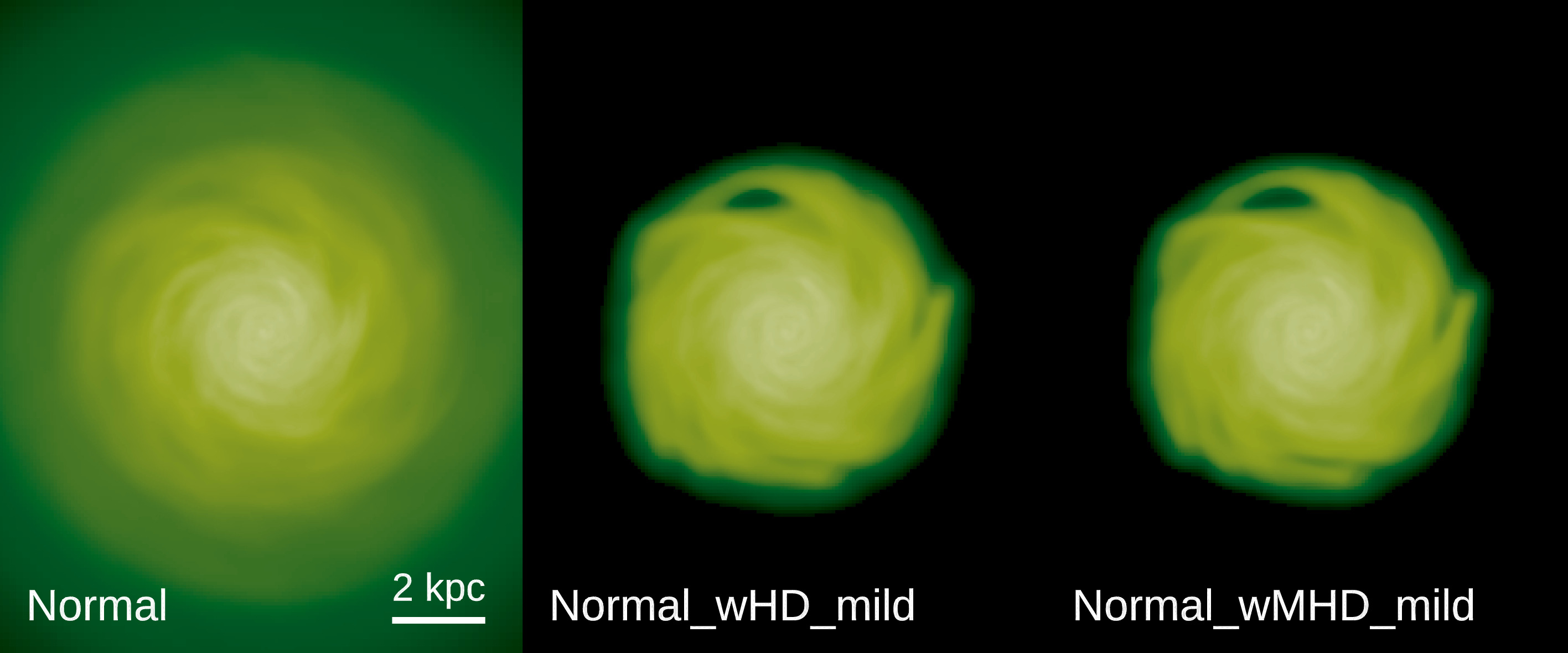}
\caption{Same as Figure~\ref{fig:disk_projection_edgeon_ncns}, however, projected in the face-on direction. Disk features are comparable with those in the \whd\ and \wmhd\ simulations.}
\label{fig:disk_projection_faceon_ncns}
\end{figure}

\begin{figure}
\centering 
\includegraphics[width=\linewidth]{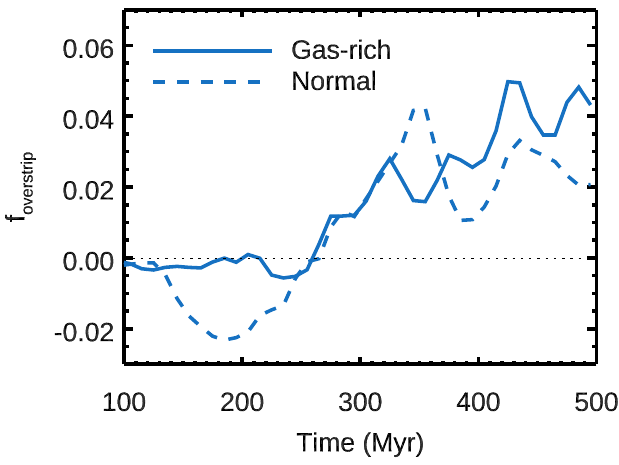}
\caption{Same as Figure~\ref{fig:overstip}, but for gas-rich (solid) and normal (dashed) disks encountering mild ram pressure. The effect of the magnetic draping layers is larger in the gas-rich disk.}
\label{fig:overstrip2}
\end{figure}


We conduct additional simulations for galaxies with no cooling and star formation to isolate the impact of cooling and stellar feedback from the results of disk gas stripping. We use the initial condition of \gbzero, however, the disk gas mass is set to $M_{\rm gas}=3.93\times10^9\,\msun$ to ensure that the galaxy has a disk gas mass similar to that of \gbzero\ immediately before encountering the ICM winds. Previous studies presenting lower gas stripping rates owing to magnetic draping layers adopt relatively mild ram pressures ($P_{\rm ram}/k_{\rm B}=6.4\times10^4{\rm cm^{-3}\,K}$ in \citealt{tonnesen14} or $P_{\rm ram}/k_{\rm B}=1.9\times10^5{\rm cm^{-3}\,K}$ in \citealt{ruszkowski14}) compared to ours ($P_{\rm ram}/k_{\rm B}=5\times10^5{\rm cm^{-3}\,K}$). Motivated by this, we hypothesize that the effect of magnetic draping layers is more apparent if the difference between ram pressure and magnetic pressure is narrowed. Therefore, we impose mild ram pressure ($P_{\rm ram}/k_{\rm B}=5\times10^{4}\,{\rm cm^{-3}\,K}$) on the galaxy along with fiducial pressure in the form of MHD ($B_x=1\,\mu G$) and HD winds.

Figures~\ref{fig:disk_projection_edgeon_ncns} and \ref{fig:disk_projection_faceon_ncns} show the hydrogen column density maps of the simulations without gas cooling and star formation. While disk and stripped clouds are highly turbulent, fragmented, and porous in star-forming galaxies (see Figure~\ref{fig:disk_projection}), gaseous disks are smooth in all cases when gas cooling and star formation are turned off. 
As shown in Figure~\ref{fig:overstip}, the disk gas is stripped less by the MHD winds than by the HD winds, and this trend is more evident when the ram pressure is weaker. However, the two cases are almost indistinguishable in appearance, which is clearly opposite to our more realistic case of cooling and star formation in Figure~\ref{fig:gas_disk}.

We also examine the effect of the magnetic draping layers on disks with different gas fractions. A lower gas fraction renders the disk less resilient to external perturbations, which can alter the effect of the magnetic draping layers. To investigate this, we set up a galaxy with $M_{\rm gas}=1.43\times10^9\,\msun$, following \citetalias{lee20}. Because the effect of the magnetic draping layers becomes more evident under mild winds (Figure~\ref{fig:overstip}), we adopt a mild ram pressure in these simulations. Figure~\ref{fig:overstrip2} shows the overstripping ratio for the gas-rich (solid) and normal (dashed) disks for 400\,Myr. As shown in Figure~\ref{fig:overstip}, the HD wind begins to overstrip the disk gas mass compared with the MHD wind runs. Overstripping is initially higher in the normal disk, but eventually becomes stronger in the gas-rich disk. This result appears to be consistent with Figure~\ref{fig:overstip} in the sense that the effect of the magnetic draping layers is better revealed in a less perturbed system.

\bibliographystyle{aasjournal}

\end{document}